\documentclass[a4paper,11pt]{article}
\usepackage[a4paper,
            bindingoffset=0.2in,
            left=1in,
            right=1in,
            top=1in,
            bottom=1in,
            footskip=.5in]{geometry}
\usepackage[T1]{fontenc}
\usepackage{inputenc}
\usepackage{amssymb} 
\usepackage{amsfonts}
\usepackage{tikz}
\usepackage{rotating}
\usepackage{makecell}
\usepackage[labelsep=endash]{caption}
\usepackage{subcaption}
\usepackage{arydshln}
\usepackage{mathtools}

\newcommand{\footremember}[2]{%
\footnote{#2}
\newcounter{#1}
\setcounter{#1}{\value{footnote}}%
}

\newcommand*{\affaddr}[1]{#1} 

\usepackage{amsmath,bm}
\usepackage{mathptmx}
\usepackage{booktabs}
\usepackage{times}

\usepackage{hyperref}
\usepackage{multirow}
\usepackage{multicol} 
\usepackage{longtable}
\usepackage{graphicx}
\usepackage{floatrow}
\usepackage{csquotes}
\usepackage[square,numbers]{natbib}
\title{On the estimation of fuzzy poverty indices}
\date{}
\author{%
  Federico Crescenzi\footremember{alley}{\affaddr{University of Tuscia. Dep. of Economics, Engineering, Society and Business administration. Viterbo, Italy. Corresponding author Email: \href{mailto:federico.crescenzi@unitus.it}{federico.crescenzi@unitus.it}}} and Lorenzo Mori \footremember{trailer} {\affaddr{University of Bologna. Dep. of . Statistical Sciences, Bologna, Italy. Email: \href{mailto:lorenzo.mori7@unibo.it}{lorenzo.mori7@unibo.it}}}
  }
\begin{document}
\maketitle
\begin{abstract}
We review the fuzzy approach to poverty measurement by comparing 
poverty indices using different membership functions proposed in 
the literature. We put our main focus on the issue of estimation
of the mean squared errors of these fuzzy methods showing which 
indices can be more accurately estimated using sample data. 
By means of simulations, we also investigate the role of parameters of the membership 
function when it comes to estimating mean squared errors via a
robustness analysis.
\end{abstract}

\section{Introduction}
\label{introduction}
Estimating poverty measures for populations of individual based on a
sample from a certain survey depends on the choice between two
paradigms: the traditional approach and the fuzzy approach. In the first
approach, an individual is regarded to as poor with respect to a poverty
predicate $(Y)$ if it belongs to the set of poor, usually defined with
a rule on $Y$. The condition of being poor or not-poor with respect to
$Y$ is therefore binary. To follow the fuzzy approach means to
allow that there is no such binary classification, instead, it
only exists individuals that are more or less deprived than others with
respect to $Y$.

It is not the objective of this paper to discuss theoretically
or philosophically about these paradigms. The differences between these
points of view are widely discussed in the books \citep{lemmi2006fuzzy};
\citep{betti2008advances}; \citep{betti2021analysis},
therefore, we would not add much more original contribution to the
debate.

It is the objective of this paper to focus on the estimation of fuzzy
poverty measures (or indices) from a statistical perspective. In fact,
to the best of our knowledge, the literature on fuzzy methods for
poverty estimation have not focused on statistical properties of this
indices either in terms of bias of the estimates nor in terms of their
mean squared errors and, ultimately, in terms of their robustness to
parameter specification. On the other hand, in a traditional approach to
poverty and inequality measurement, the diffusion of sampling variances
and assessments of indices has been widely addressed in the
literature (see \citep{kakwani1993statistical};\citep{graf2014variance};\citep{duclos2006robust}). Also,
R-package Laeken offers routines to estimate the variance via a
non-parametric bootstrap approach (\citep{alfons2012estimation}).

The comparison between these methods is not straightforward because each
fuzzy poverty index depends on the so-called membership function that
ranks individuals in terms of their deprivation with respect to the
poverty predicate. This function represents the researcher's belief on
the poverty phenomenon and therefore different functions are
intrinsically difficult to be compared. Moreover, they often depend on
different parameters and usually authors suggest different synthetic
measures over the function. Our comparison tackles this issue by
addressing the mean squared error of these measures. In fact, while a
approach may be appealing in theory, ultimately, when it comes to
produce estimates from samples it is necessary to have estimates with
good properties. To do so however, we need a common sample statistic to consider. We
decided for the expected value of the membership function as it is the
statistic that is reported by many (but not all) fuzzy methods, has
known statistical properties, and it is applicable to all membership
functions. For example, the axiomatic approach lacks of axioms on the
mean squared error of the estimator which may be important for small
samples or sub-populations.

We address all these topics by setting out a simulation study
that consists in creating a synthetic population from a real sample of
individuals and evaluating the bias, the bias in the mean squared error,
the coefficient of variation, and the robustness of the methods
simulating two different sample surveys. The first one uses simple
random sampling of individuals, the second one uses a more complex
sampling design. All the measures discussed are compared using bootstrap
and jackknife replications. In addition, for those indices that
require the researcher to define parameters, we check their robustness
and the impact, in reliability terms, at as these values change.

The remainder of the paper is so structured.
Section~\ref{sec-a-brief-history-of-poverty-measurement} offers a brief
history of poverty measurement for the not experienced reader although it has
not to be intended as a comprehensive literature review of poverty
estimation. Section~\ref{sec-membership-functions} discusses in analytic
details the fuzzy membership functions. Section~\ref{sec-simulation-set-up-data-mean-squared-errors-and-sperformace-measures.} introduces the
simulation set-up and revises the bootstrap and the jackknife approach
that is used in the results Section~\ref{sec-results} . Note that the
jackknife procedure that we use here is an ad-hoc procedure for fuzzy
poverty estimates suggested by \citep{betti2018simplified}. Section ~\ref{sec-robusteness} discuss the robustness of the fuzzy indicators. 
Eventually, Section~\ref{sec-conclusions} concludes the work and gives
insights for further research.

\section{A brief history of poverty 
measurement}\label{sec-a-brief-history-of-poverty-measurement}

From a statistical point of view the main problem to evaluate poverty is the definition of a certain threshold beyond which a statistical unit
(i.e. an individual or household) is defined poor. This threshold is usually a function of a (monetary) variable
denoting the welfare status of the unit. The definition of the poverty line is not only a statistical problem but it is also political and in general, from an official statistics perspective, it is held fixed by national statistical institutes. It is customary to trace the origin of the poverty line back to the works published in the 1960s by Mollie Orshansky, but this concept places its origins well before (\citep{fisher1997development}). 
In order to understand the needs that led researchers to define fuzzy poverty measures, we review concisely the history of the poverty line (\citep{fisher1992development} and \citep{allen2013poverty}).

 In 1887 Charles Booth during a meeting of the Royal Statistical Society presents the results of a survey conducted in London. In this presentation, for the very first time, Booth uses the concept of "line of poverty". A line defined in terms of daily income that admits to divided the population in to two sub-set the poor and the non-poor one (\citep{History1}).

During the 20th century three different concept of  poverty were developed: subsistence, basic needs and relative
deprivation. The concept of subsistence - which spread in Great Britain - was the first to be developed by  Benjamin Seebohm Rowntree in 1901. Rowntree  set a poverty line in terms of a minimum weekly sum of money for a healthy life. This paradigm defines as poor those who are unable to obtain the minimums resources to ensure an efficient physical condition and was created in cooperation with nutritionists under the responsibility of entrepreneurs (\citep{Rowntree}).

In 1964 were published the recalled works of Mollie Orshansky who defined the poverty threshold as a measure of income inadequacy by taking the cost of food plan per family. This threshold is officially adopted, in 1969 by the American inter agency poverty level review committee. All these three approaches, Booth, Rowntree and Orshansky, lead to the concept of absolute poverty. Within this approach the threshold is fixed to a monetary quantity (usually a daily one) necessary to reach the basic life necessities. The absolute poverty approach is still used, especially in the under-developed countries, and the World Bank use a threshold of $1.90\$$ per day.

Few years later, in 1976, a new formulation to express the concept of poverty, based on basic needs, was introduced at the International Labour Organization's World Employment Conference (\citep{jolly1976world}). This approach enlarged the concept of subsistence with the feature of circumscribing poverty in a context of development social and economic status of a country. It is still used for the developing countries. 

Following the ideas of Adam Smith the concept of a relative poverty become to be more and more popular from the 1950s. The idea that is not possible to define a threshold of poverty without considering the community in which individuals live find bring, in 1967, Victor Fuchs to define the poverty threshold as one-half the median family income. From 1967 on-wards this concept remain almost stable with only small changes in the value of the percentage of the median family income to be considered. Today the most common used threshold is set at $60\%$ of the median household income. Relative poverty measures are officially used, among others, by the European Union, UNICEF and the OECD. 


The dichtomization of the population in two non-overlapped subsets of
poor units and not-poor units was criticized in the early 90s of the
last century (\citep{chelilemmi}). At the core of the debate that criticizes the
poor/not-poor dichotomy there is the fuzzy approach. The aim is to
obtain index able to capture a complicated social phenomenon that cannot
be reduced to a trite true/false (poor/non-poor) binary logic. This
paradigm replaces the dichotomy with a measure of degree (or propensity) to be poor defined as a function of the the distribution of income (or an equivalent poverty
predicate).

This propensity is based on the fuzzy set approach (\citep{zadeh1965fuzzy};
\citep{klir1995fuzzy}) in which a unit $i$ does not belong or not belong
to set of poors $A$ (i.e either $i\in A$ or $i \notin A$) but it
is given a propensity or degree of poverty or deprivation with respect
to the poverty predicate.


The function that maps each unit to the fuzzy set approach is called the
(fuzzy) membership function. Interestingly, this approach comprehends
the traditional approach as this can be viewed as a fuzzy approach that
uses a staircase-membership.


\section{Fuzzy approaches to poverty estimation}
\label{sec-membership-functions}

The traditional approach to poverty measurement instead is based on the
crisp sets theory, where the population is divided into the subset of
poor and the subset of non-poor according to a rule on $\mathcal{Y}$.
For example people whose $Y$ level falls below the poverty line $\tau$ are regarded as poor (1) while the remaining as not-poor (0). We thus say that poverty
measurement based on crisp set assigns a value of either 1 or 0 to each
individual in the universal set according to the rule $(1 (y_i < \tau))$.

In the fuzzy approach, given a poverty predicate $Y$ (e.g. the equivalised disposable income), its support
$\mathcal{Y}$, 
and the (fuzzy) set $\mathcal{A}$, a membership function $\mu$ is a
mapping between the support of $Y$ to
$\mathcal{A} =[0,1] \subset \mathbb{R}$,
i.e.~$\mu: \mathcal{Y} \to \mathcal{A}$. The higher is the value of $\mu$ the
most is the membership of the unit to the set. Each fuzzy set is uniquely and
completely defined by its membership function. 

Two of the most important
concept of fuzzy sets, used also in fuzzy index, are the $\alpha-cut$
and the strong $\alpha-cut$. Given a fuzzy set $\mathcal{A}$ defined
on $Y$ and any number $\alpha \in [0,1]$, the $\alpha-cut$
($^{\alpha}\mathcal{A}$) and the strong $\alpha-cut$
($^{+\alpha}\mathcal{A}$)  are the crisp sets
$^{\alpha}\mathcal{A}=\{y|\mathcal{A}(y)\ge \alpha\}$ and
$^{+\alpha}\mathcal{A}=\{y|\mathcal{A}(y)>\alpha\}$. That is, the $\alpha-cut$ of a fuzzy set $\mathcal{A}$ is the crisp
set $^{\alpha}\mathcal{A}$ that contains all the elements of the
universal set $Y$ whose membership grades in $\mathcal{A}$ are
greater than or equal to the specified value of $\alpha$. The concept
of the strong $\alpha-cut$ is the same with value that can be only
greater than the specified value of $\alpha$. A last definition that
will return  in the
following is the cardinality of a fuzzy set,
i.e. $Card(A)=|A|=\sum_i\mu(y_i)$.

A first roughly classification based on the type of membership function
assumed distinguishes between distance based and distribution based
membership functions.

\hypertarget{distance-based-membership-functions}{%
\subsection{Distance based membership
functions}\label{distance-based-membership-functions}}

A number of authors developed fuzzy indices on the basis of the
concept of $\alpha-cut$. In this case, the membership function is defined through a
distance between $y$ and a threshold $z$ $(y,z\in\mathcal{Y})$ of
the kind of

\[
\mu(y_i)=\begin{cases}
1, \quad 0<y_i<z_1\\
f(y_i), \quad z_1\le y_i<z_2\\
0, \quad y_i\ge z_2
\end{cases}
\]

where $f(y_i)$ is a positive decreasing function. The literature have suggested
several membership function  depending on the definition of the
function $f(\cdot)$. According to \citep{martinetti2006capability} the distance
based membership functions can be classified into triangular, trapezoidal and Gaussian functions. In triangular functions, $f(\cdot)$ is expressed as the combinations of two different functions \(f_1(\cdot)\) and \(f_2(\cdot)\). In trapezoidal unctions, $f(\cdot)$ is of the form $\frac{z_2-y_i}{z_2-z_1}$, with \(z_1\) and \(z_2\) being two threshold values. A special case of
  the trapezoidal are the linear functions where \(z_1\) and \(z_2\)
  are, taken as the maximum and minimum values of \(Y\). For Gaussian functions $f(\cdot)$ is expressed in the form of a Gaussian distribution.
At the best of our knowledge, there is no Gaussian functions for fuzzy
poverty indicators.

In chronological order, the most used distance based functions are \citep{cerioli1990fuzzyart}, \citep{belhadj2011new}, \citep{zedini2015new}, \citep{besma2014employment}, and Chakravarty (2019).

\subsubsection{Cerioli and Zani (1990)}

The first proposal of a membership function is that of \citep{cerioli1990fuzzyart} who suggest

\begin{equation}\protect\hypertarget{eq-cerioli}{}{ 
\mu(y_i)=\begin{cases}
1, \quad 0<y_i\le z_1\\
\frac{z_2-y_i}{z_2-z_1}, \quad z_1< y_i<z_2\\
0, \quad y_i\ge z_2
\end{cases}
}\label{eq-cerioli}\end{equation}

This formulation states that the membership function is a simple trapezoidal form leading
to a fuzzy index in which the values of $z_1$ and $z_2$ have to be chosen by the researcher, and a linear function
\(f(\cdot)\) is used for all intermediate values.

\subsubsection{Belhadj  (2011)}

\citep{belhadj2011new} suggests the following membership function

\begin{equation}\protect\hypertarget{eq-belhadj2011}{}{\mu(y_i) = \begin{cases}
        1                                                                       & 0 < y_i < z_{min} \\
        \frac{-x_i}{z_{\max} - z_{\min}} + \frac{z_{\max}}{z_{\max} - z_{\min}}        & z_{min} \le y_i < z_{max}\\
        0                                                                       & y_i \ge z_{max}
    \end{cases}
}\label{eq-belhadj2011}\end{equation}

where \(z_{\min}, z_{\max}\) are defined respectively as the upper and
lower poverty line. The upper poverty line corresponds to the level of
the total expenditure per-capita necessary so that the households can
satisfy their basic food needs without sacrifice while the lower poverty
line includes minimum expenditure to satisfy basic food and non-food
needs. Remarkably, we notice that this membership function is a special
case of Equation~\ref{eq-cerioli} for given values of the threshold
parameters. Therefore, throughout the paper we will not make any further
distinction between Equation~\ref{eq-cerioli} and
Equation~\ref{eq-belhadj2011}.
\subsubsection{Zedini and Belhadj (2015) }

\citep{zedini2015new} propose another triangular membership function defined by:

\begin{equation}\protect\hypertarget{eq-ZBM}{}{
\mu(y_i) = \begin{cases}
        1                                   &   a \le y_i < b\\
        \frac{-y_i}{c-b} + \frac{c}{c-b}    &   b \le y_i < c\\
        0                                   &   x_i < a \cup y_i \ge c\\
    \end{cases}
}\label{eq-ZBM}\end{equation}

where \(a,c,b\) are percentiles estimated with via the bootstrap  over a series of fuzzy latent poverty states (see \citep{zedini2015new} for more details).

\subsubsection{Belhadj (2014)}\label{besma2014}

\citep{besma2014employment} suggest the following trapezoidal membership function

\begin{equation}\protect\hypertarget{eq-belhadj2015}{}{
\mu(y_i) = \begin{cases}
        1                                                                           & y_i < z_1\\
        \mu_1 = 1-\frac{1}{2}\left(\frac{y_i-z_1}{z_1}\right)^\beta             & z_1 \le y_i < z^*\\
        \mu_2 = 1-\frac{1}{2}\left(\frac{z_2 - y_i}{z_2}\right)^\beta         & z^* \le y_i < z_2\\
        0                                                                           & y_i \ge z_2
    \end{cases}
}\label{eq-belhadj2015}\end{equation}

where \(z^{*}\) is the flex point of \(\mu\) and \(\beta\) is a shape
parameter ruling the degree of convexity of the function. In particular,
when \(\beta=1\) the trend is linear.

\subsubsection{Chakravarty (2019)}\label{chakravarty2019}

Finally, Chakravarty (2019) presents five important properties that
should be fulfilled by a membership function~when defining a fuzzy
poverty index that are Homogeneity of Degree Zero, Linear
Decreasingness, Continuity, Maximality, and Independence of Non-meager
Attribute Quantities. On the basis of such axioms the author suggests
the following triangular membership function:

\begin{equation}\protect\hypertarget{eq-chackra}{}{
\mu(y_i) = \begin{cases}
        1                           & y_i = 0\\
        \frac{z_2 - y_i}{z_2}       & 0 \le y_i < z_2\\
        0                           & y_i \ge z_2
    \end{cases}
}\label{eq-chackra}\end{equation}

Note as, again, the membership function defined by Chakravarty (2019) can be viewed as a
special case of the Equation~\ref{eq-cerioli} for \(z_1=0\).

\hypertarget{distribution-based-membership-functions}{%
\subsection{Distribution based membership
functions}\label{distribution-based-membership-functions}}

\subsubsection{Cheli and Lemmi (1995)}\label{cheli1995totally}

Alongside to the distance based membership function there are the distribution based membership function
at which belongs the indicators defined by \citep{chelilemmi}. The authors proposed the ``Totally Fuzzy and Relative'' (TFR) approach
defining the membership function for a monetary variable as:

\begin{equation}\protect\hypertarget{eq-TFR}{}{
\mu(y_i)=\left(1-F_Y(y_i)\right)^{\alpha}
}\label{eq-TFR}\end{equation}

where \(F_Y\) is the cumulative distribution function of \(Y\)
calculated for the \(i\)-th individual. The parameter \(\alpha\ge 1\) is
chosen so that the average fuzzy measure equals the traditional head
count ratio for the overall population.

\subsubsection{Betti and Verma (1999)}
\label{betti1999}

Starting from TFR approach, \citep{betti1999measuring} suggested to replace
the cumulative distribution function of $(Y)$ with the Lorenz curve
$(L)$ obtaining

\begin{equation}\protect\hypertarget{eq-betti-verma}{}{
\mu(y_i)=(1-L_Y(y_i))^{\alpha-1} = \left(\frac{\sum_j w_jy_j|y_j> y_i}{\sum_j w_jy_j|y_j> y_1}\right)^{\alpha-1}
}\label{eq-betti-verma}\end{equation}

The parameter \(\alpha\ge 1\), as before, is chosen so that the average
fuzzy measure equals the traditional head count ratio for the overall
population.

\subsubsection{Betti et al. (2006)}
\label{betti2006multidimensional}

Based on the TFR approach, \citep{betti2006multidimensional} take into account both
the proportion of individuals less poor than the person concerned,
$(1-F_Y(y_i))$, and the share of the total equivalised income received
by all individuals less poor than the person concerned to propose the
following membership function

\begin{equation}\protect\hypertarget{eq-betti2006}{}{
\begin{split}
     \mu(y_i)&=\left(1-F_Y(y_i)\right)^{\alpha-1}\left(1-L_Y(y_i)\right)=\\
    &=\left(\frac{\sum_j w_j|y_j> y_i}{\sum_j w_j|y_j> y_1}\right)^{\alpha-1} \left(\frac{\sum_j w_jy_j|y_j> y_i}{\sum_j w_jy_j|y_j> y_1}\right)
\end{split}
}\label{eq-betti2006}\end{equation}

where $w_i$ is the sampling weight of statistical unit $i$ and
$\alpha$ is computed as before. 

\subsection{A further distinction between membership function: the
economical
prospective}
\label{a-further-distinction-between-membership-function-the-economical-prospective}

A well-known classical economical classification divided index in two
large sets of positive and normative indices. Positive indices are those
ones with a completely objective definition (e.g. Theil index, Gini
index...) On the other hand, indices are defined as normative if their definition is somewhere subjective (e.g.  Atkinson index). This distinction, in our view, can be used for fuzzy
indices, as well. As seen so far every single membership function has at least one
parameters to be defined but not all of them have a mathematical
algorithm to be defined and have to be chosen by the researcher.

\paragraph{Positive indices}

  Equation~\ref{eq-cerioli} in the special case in which
  $(z_1=min(y_i))$ and $(z_2=max(y_i))$. Equation~\ref{eq-TFR}, Equation~\ref{eq-betti-verma}, and
  Equation~\ref{eq-betti2006} in all those cases the
  parameter $(\alpha)$ is chosen according so that the
  expected value of the membership function equals the head count ratio (i.e. the proportion of poors).
Equation~\ref{eq-ZBM} as noted before the Zedini and Belhadj (2015) membership function have three
  different parameters a,b and c to be estimated through bootstrap
  techniques. Authors suggests to create a set of 100 membership function corresponding
  to 100 fuzzy sets. The first membership function will have a=0, $b=\widehat{p}_1$ and
  $c=\widehat{p}_2$; where $\widehat{p}_1$ and $\widehat{p}_2$
  are, respectively, the bootstrap estimates of the first and second
  percentiles of the poverty predicate variable. For the membership function from 2 to 99
  they suggest to use $a=\widehat{p}_{j-1}$, $b=\widehat{p}_j$ and
  $c=\widehat{p}_{j+1}$ with $j=2,\dots,99$. The last membership function for
  $j=100$ is defined with $a=\widehat{p}_{99}$,
  $b=0.5(\widehat{p}_{99}+\widehat{p}_{100})$ and
  $c=\widehat{p}_{100}$ . The final membership function is then obtained on the
  basis of the cardinality (presented in the previous section) approach and the divisive hierarchical algorithm.

\paragraph{Normative indices}
All the others (Equation~\ref{eq-cerioli},
Equation~\ref{eq-belhadj2015} and Equation~\ref{eq-chackra}) are
normative indices and, in our view only the index
Equation~\ref{eq-belhadj2011} deserve for further attention. Authors, in
this case suggests to set the values of $(z_{min})$ equal to:

\[
z_{min}=(2-\alpha)z^f
\] where $\alpha$ is the expected non-food shares of households,
$z^f$ is the food poverty line while the value of $z_{max}$
correspond to the level of the total consumption per capita necessary so
that the households can satisfy their basic food needs without
sacrifice. This poverty line can be obtained by an iterative algorithm.
Note as \citep{belhadj2011new} allow the use of $z_{min}$ and $z_{max}$
specific by area. However, those thresholds require to the researcher to
define the value for which households can satisfy their basic food needs
without sacrifice. For this reasons we prefer to consider this index as
normative index. In addition, given that the the membership function is defined exactly
as for Equation~\ref{eq-cerioli} and that for this index we analyze a
number of different thresholds we consider the two indices as the
same one.

\section{Simulation set-up: mean squared errors estimators,  data, and measures of performance.}
\label{sec-simulation-set-up-data-mean-squared-errors-and-sperformace-measures.}

In order to estimate the mean squared error of a sample statistic the jackknife repeated replications and the bootstrap are two widely used
tools in sample surveys. The main difference between those two is that
the first one is definable only if all the sample information are
available, e.g.~primary selection unity and strata, while the second one
is a simplified version which doesn't necessarily need the sampling information.

\subsection{Non-parametric Bootstrap}

The bootstrap is probably the most known re-sampling method to
estimate the variance of a statistic. There are many extensions and variants of the bootstrap,
which can be divide into parametric and non-parametric bootstrap. We focus on the
non-parametric version of the bootstrap, which is the one that it is also provided with the
excellent R-package Laeken (\citep{alfons2012estimation}). The procedure consists in sampling
with replacement R  samples of size M, then, for each bootstrap sample calculate a given
statistic $S_r$. The variance of the bootstrap replicates is then used as an estimate of the
unknown variance of the statistic of interest. The bootstrap variance estimator is based on the idea that the inference about a population from sample data an be modeled by re-sampling
the sample data and performing inference about a sample from re-sampled data.

\subsection{Jackknife Repeated
Replications}
\label{jackknife-repeated-replications}

Jackknife repeated replication provides a versatile and straightforward
computational technique for variance estimation. Sometimes, the survey
may be designed so that there exists information on strata, primary
sampling units, and computational units. In this situation, the
jackknife approach may be an invaluable tool to estimate the variance of
a given sampling statistic.

\citep{betti2018simplified} have extended this method for
estimating variances for sub-populations (including regions and other
geographical domains), longitudinal measures - such as persistent
poverty rates and measures of net changes - and averages over
cross-sections in rotational panel designs and also to fuzzy poverty
measures. In the standard delete one-primary
sampling unit (PSU) at a time Jackknife' version, each replicate is
formed by eliminating one sample PSU from a particular stratum at a time
and increasing the weight of the remaining sample PSU's in that stratum
appropriately so as to obtain an alternative but equally valid estimate
to that obtained from the full sample. Briefly, the standard jackknifing
involves the following. Let $S$ be a full-sample estimate of any
complexity. We use the subscript $i$ to indicate a sample PSU and
$h$ to indicate its stratum; $a \ge 2$ is the number of PSUs in
stratum $h$.

Let $S_{(hi)}$ the estimate produced using the same procedure after
eliminating primary unit $i$ in stratum $h$ and increasing the weight of
the remaining $a_h-1$ units in the stratum by an appropriate factor
$g_h$. Let $S_{(h)}$ be the simple average of the $S_{(hi)}$ over
the $a_h$ sample units in $h$. The variance of $S$ is then
estimated as:

\[
var(S)=\sum_h\biggl[(1-f_h)\times \sum_i g_{(hi)}(S_{(hi)}-S_{(h)})^2\biggr]
\]

where $(1-f_h)$ is the finite population correction, and it is
approximately equal to 1 for samples in typical social surveys. While
one may take factor $g_{(hi)}$ to be independent of the particular i
in a given stratum $h$, \citep{verma2011taylor} propose to use:

\[
g_{(hi)}=\frac{a_h}{(a_h-1)}
\]

where $w_h=\sum_i w_{hi}$, with $w_{hi}=\sum_j w_{hij}$ as the sum
of sample weights of ultimate units $j$ in primary selection units
$i$. This means that in each replication $h_i$, the weights for
individual units are redefined and re-scaled as follows:

\[
w^1_{hij} = \begin{cases}
w^1_{hij}=w_{hij}                     & j \notin h \\
w^1_{hij}=g_{(hi)}\times w^1_{hij}    & j \in h, j \notin i \\
w^1_{hij}=0                           & j \in h, j \in i
\end{cases}
\]

$g_{(hi)}$ is in line with the jackknife variance estimation for unequal
probability sampling proposed by \citep{berger2005jackknife}. \citep{betti2018simplified} introduce the
following factor: $\biggl(1-\frac{w_{hi}}{w_h}\biggr)$ as a correction
factor for unequal probabilities, `reducing the contribution of
observations which have higher $\pi_i$ values (selection
probabilities) and thus make smaller contributions to variance'. The
inverse of this correction to selection probability is the corresponding
correction to weight $g_{hi}$.

\subsection{Data and measures of performance}
In order to compare all the fuzzy approaches introduced in
Section~\ref{sec-membership-functions}, we explore two different
scenarios by means of two different simulations. The two simulations
share the same population $\mathcal{P}$. In order to avoid as much as possible the
results to be dependent from authors' source code, the population that
we use is simulated using the R package simPop (\citep{templ2017simulation}). We
remind the reader to the package for full details on the simulation
process although, in brief, the procedure uses a real sample of
individuals\footnote{The package uses an anonymised version of the
  Austrian EU-SILC survey (\citep{alfons2012estimation})} and
reconstructs a synthetic population of individuals that have generated
that sample. It starts by applying iterative proportional fitting
(\citep{deming1940least}) to obtain a calibrated survey sample from the
observed sample, then trough replications and statistical models for
categorical variables it estimates a joint model of the variables in the
survey and eventually it generates a synthetic population using
simulated annealing. Once that we have the synthetic monetary measure in
the population (i.e.~the equivalised disposable income) we set up the
two different scenarios. 

The first scenario is that of a survey where the
  sample is selected following a simple random sampling without
  replacement scheme. This scenario is denominated as ``SRS''. The second scenario that we investigate is that of a more complex
  survey design that involves the definition of strata and primary
  sampling units mimic the usual sampling scheme used by Eurostat for
  the EU-SILC or HBS. As strata we use the NUTS 2 area while as PSUs the
  households ID. This scenario is denominates as ``Complex''. 
  
Table \ref{tab0} reports the sample size ($n$) for each area in the two scenarios and the size of the population ($N$). The sampling fraction is approximately 1\% for each of the two scenarios

\begin{table}[H]
\caption{Sampled units under the two scenarios and corresponding population units}\tabularnewline

\centering
\begin{tabular}{lrrr}
\hline
Area & $n_{SRS}$ & $n_{Complex}$ & $N$\\
\hline
Burgenland & 33 & 6 & 2905\\
Carinthia & 77 & 55 & 5546\\
Lower Austria & 176 & 157 & 16232\\
Salzburg & 113 & 124 & 14262\\
Styria & 43 & 48 & 5344\\
Tyrol & 110 & 128 & 12107\\
Upper Austria & 65 & 100 & 7219\\
Vienna & 162 & 187 & 17686\\
Vorarlberg & 42 & 3 & 3756\\
\hline
Total & 821 & 808 & 85057\\
\hline
\end{tabular}
\label{tab0}
\end{table}
\noindent
We calculate for each individual in each sample his membership function
according to the different functions that we outlined in
Section~\ref{sec-membership-functions}. For the membership functions
that require parameters to be specified we set arbitrarily (and without
loss of generality) the parameters to some quantiles of the population
distribution (see Table \ref{tabparam}). 

\begin{table}[H]
\caption{Membership functions and parameter settings in synthetic population (
$Y_p$ denotes the $p$-th quantile of the empirical distribution of $Y$).}
\centering
\begin{tabular}{lcl}
\hline
Function & Parameters & Values\\
\hline
Belhadj (2015) & \makecell[c]{$z_1$\\$z_2$\\$\beta$} & \makecell[l]{$Y_{0.01}$\\$Y_{0.99}$\\$2$}\\
\hline
Cerioli and Zani (1990) & \makecell[c]{$z_1$\\$z_2$} & \makecell[c]{$Y_{0.001}$\\$Y_{0.99}$}\\
\hline
Chakravarty (2019) & $z_2$ & $Y_{0.50}$\\
\hline
Cheli and Lemmi (1995) & $\alpha$ & $\sum_{i\in \mathcal{P}} 1 (y_i\le \tau)$\\
\hline
Betti and Verma (1999) & $\alpha$ & $\sum_{i\in \mathcal{P}} 1 (y_i\le \tau)$\\
\hline
Betti et. al., (2006) & $\alpha$ & $\sum_{i\in \mathcal{P}} 1 (y_i\le \tau)$\\
\hline
\end{tabular}
\label{tabparam}

\end{table}
\noindent
A key fact that is important to remark here is that the comparison over
all indices is done taking the average of the membership functions over
the population as the true value. Therefore, the corresponding sample
statistic that we consider is

\begin{equation}\label{eq-statistic} \hat H = \sum_{i=1}^n\mu_i\times w_i \end{equation}

However, it is often the case that rather than estimating a value for
the whole population, the interest is in producing estimates for given
sub-domains\footnote{Sub-domains need not necessarily be geographical
  areas, even though it is was we explore in this paper.}. Therefore,
Let $H_{j^*}$ be the true area parameter in the $j^*$-th area and
$\widehat{H}_{j^*t}$ its estimate in replicate $(t=1,\dots,T,$ with  $T=500)$ using one
of the previous defined fuzzy membership functions.

We evaluate each fuzzy approach in terms of bias, mean squared error and
coefficient of variation. Let us start from the bias of the estimates.
For each area the bias is defined as:

\[ Bias_{j^*} =\frac{1}{T} \sum_{t=1}^T (\widehat{H}_{j^*t}-H_{j^*}) \]
To evaluate the reliability of the estimates we use the Coefficient of
Variation (CV) and the Coefficient of Variation of the Second order
($CV2$), a robustified version of CV (\citep{kvaalseth2017coefficient}) which relay on
the $([0,1])$ interval. For the generic area $j^*$ they are defined as:

\begin{align}
    CV_{j^*}  &= \frac{1}{T}\sum_{t=1}^T\frac{\sqrt{\widehat{MSE}(\widehat{H}_{j^*t})}}{\widehat{H}_{j^*t}} \\
    CV2_{j^*} &= \sqrt{\Biggl(\frac{CV_{j^*}^2}{1+CV_{j^*}^2}\Biggr)} 
\end{align}

where $\widehat{MSE}(\widehat{H}_{j^*t})$ is the estimated Mean Square
Error at the replicate $t$ for the area $j^*$. 

To evaluate the two variance procedures we also compare: Average True Mean Square Error
(ATMSE), Average Estimated Mean Square Error (AEMSE) reporting also the
bias of the MSE (BMSE)

\begin{align}
    ATMSE &= \frac{1}{J^*}\sum_{j^*=1}^{J^*}\biggl\{\frac{1}{T}\sum_{t=1}^T\biggl(\widehat{H}_{j^*t}-H_{j^*t}\biggr)^2\biggr\} \\
    AEMSE &=\frac{1}{J^*}\sum_{j^*=1}^{J^*}\biggl\{\frac{1}{T}\sum_{t=1}^T\widehat{MSE}(\widehat{H}_{j^*t})\biggr\}\\
     BMSE &=\frac{1}{J^*}\sum_{j^*=1}^{J^*}\Biggl\{\frac{1}{T}\sum_{t=1}^T(\widehat{MSE}(\widehat{H}_{j^*t})-MSE(H_{j^*t}))\Biggr\}
\end{align}

where $MSE(H_{j^*t})$ is the true Mean Square Error at the replicate $t$.

\section{Results}\label{sec-results}

This section reports the results from our simulations separately for a
simple random sampling of units of the synthetic population and for a
complex sampling of the same units. For the first case, we report only
estimates of mean squared error that use the bootstrap as the jackknife
is designed to work with more complex designs.

\subsection{Simple random sampling
design}\label{simple-random-sampling-design}

Table~\ref{tab1} reports the values of bias, CV and CV2 for all the
areas and at national level for all the indices.  We notice
that regardless of the membership function considered each estimator is
unbiased, or at least, the amount of the estimated bias is negligible. This is not a surprising results as we are considering estimator of the
kind seen in Equation~\ref{eq-statistic}. Being unbiased estimators, we
may proceed on by discussing the variance rather than using the more
comprehensive definition of mean squared error. However, to keep
consistency we shall keep on referring to the mean squared error.

\begin{figure}[H]

\begin{minipage}[t]{.3\textwidth}

{\centering 

\raisebox{-\height}{

\includegraphics[width=\textwidth]{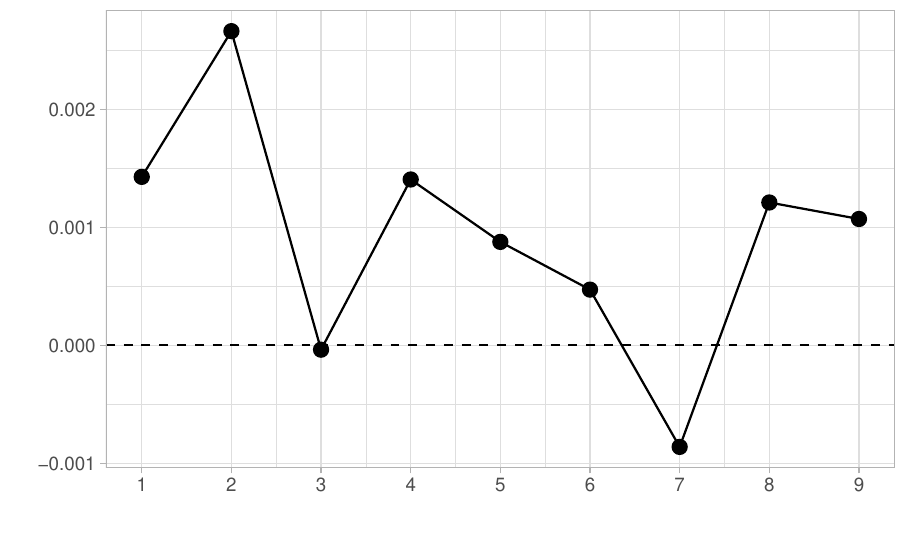}

}

}

\subcaption{\label{fig-BiasSimpleAreas-1}Belhadj (2014)}
\end{minipage}%
\begin{minipage}[t]{.3\textwidth}

{\centering 

\raisebox{-\height}{

\includegraphics[width=\textwidth]{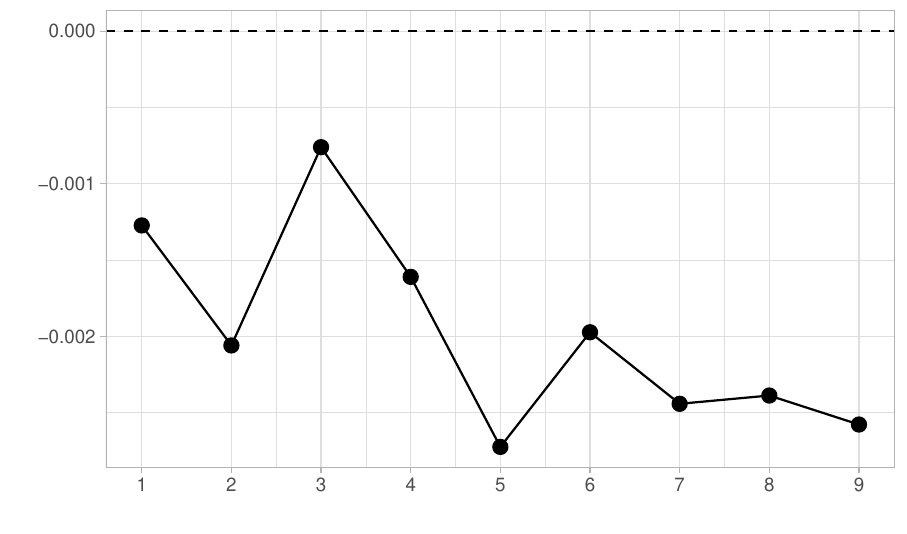}

}

}

\subcaption{\label{fig-BiasSimpleAreas-2}Cerioli and Zani (1999)}
\end{minipage}%
\begin{minipage}[t]{.3\textwidth}

{\centering 

\raisebox{-\height}{

\includegraphics[width=\textwidth]{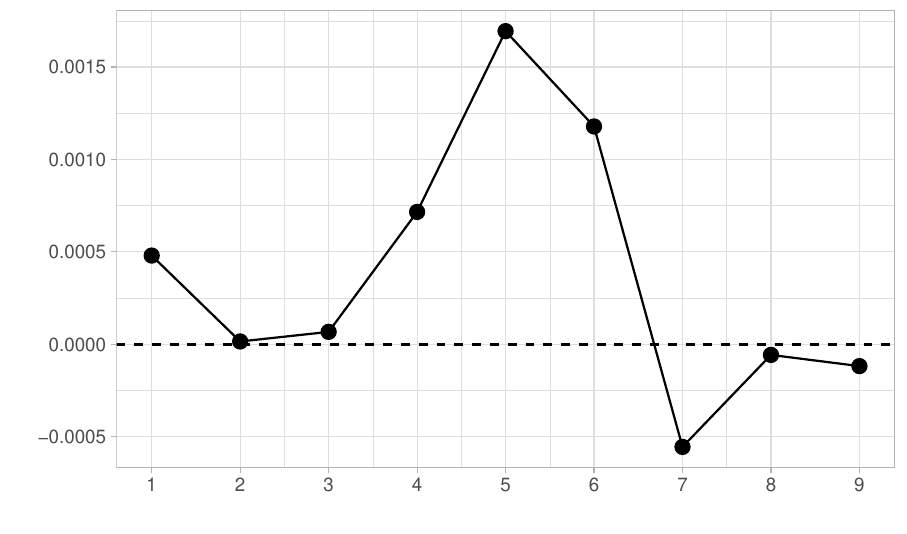}

}

}

\subcaption{\label{fig-BiasSimpleAreas-3}Chackravarty (2019)}
\end{minipage}%
\newline
\begin{minipage}[t]{.3\textwidth}

{\centering 

\raisebox{-\height}{

\includegraphics[width=\textwidth]{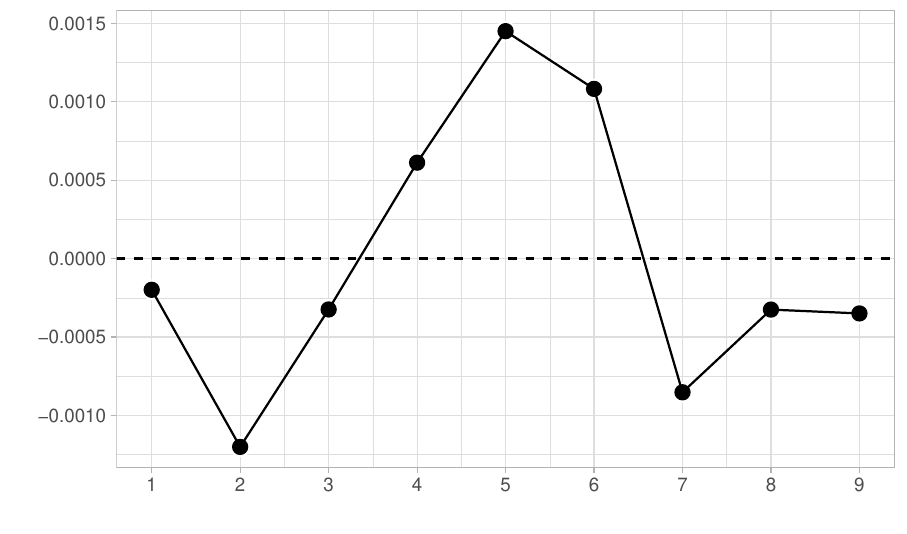}

}

}

\subcaption{\label{fig-BiasSimpleAreas-4}Cheli and Lemmi (1995)}
\end{minipage}%
\begin{minipage}[t]{.3\textwidth}

{\centering 

\raisebox{-\height}{

\includegraphics[width=\textwidth]{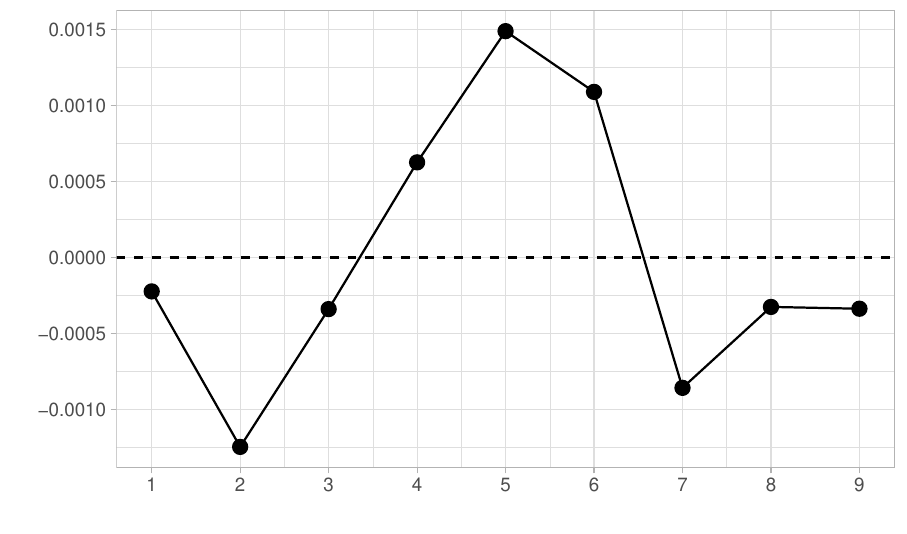}

}

}

\subcaption{\label{fig-BiasSimpleAreas-5}Betti and Verma (1999)}
\end{minipage}%
\begin{minipage}[t]{.3\textwidth}

{\centering 

\raisebox{-\height}{

\includegraphics[width=\textwidth]{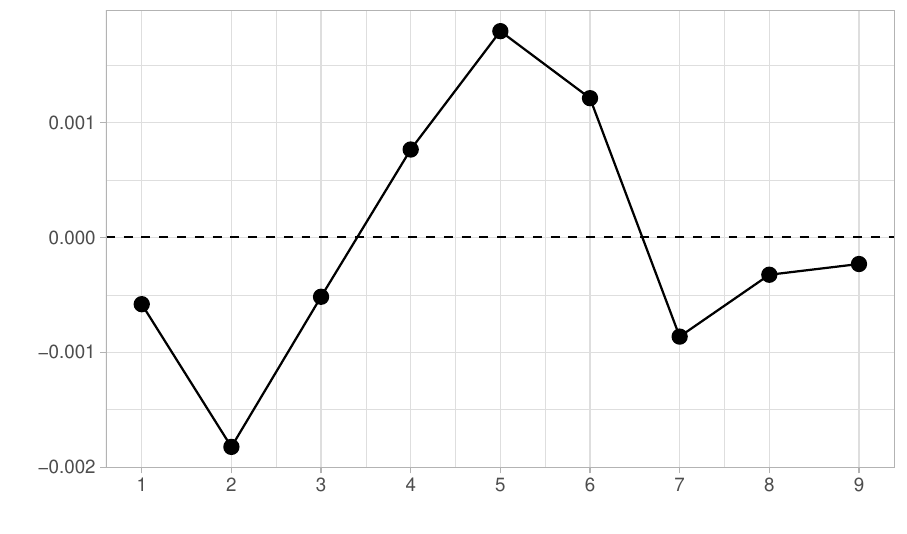}

}

}

\subcaption{\label{fig-BiasSimpleAreas-6}Betti et. al (2006)}
\end{minipage}
\newline

\begin{minipage}[t]{.3\textwidth}

{\centering 

\raisebox{-\height}{

\includegraphics[width=\textwidth]{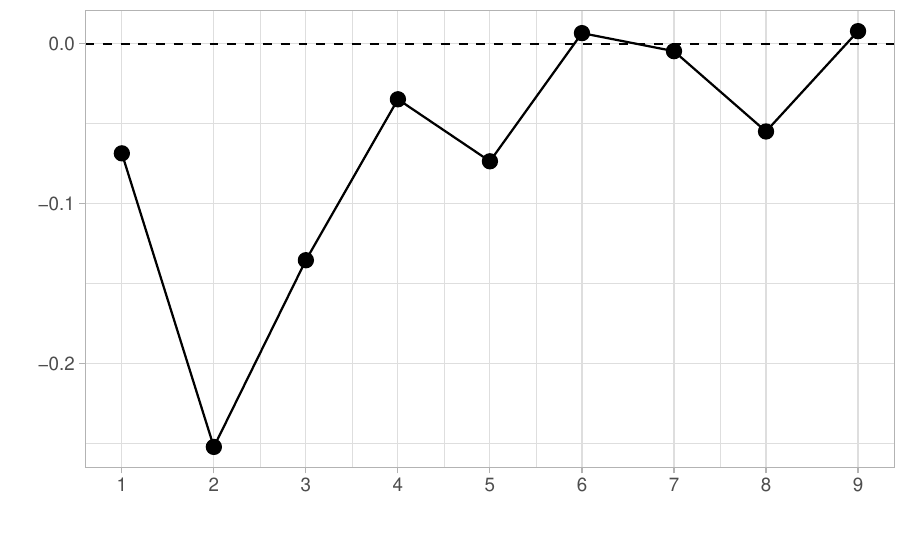}

}

}

\subcaption{\label{fig-BiasSimpleAreas-ZBM}Zedini and Belhadj (2015)}
\end{minipage}%

\caption{\label{fig-BiasSimpleAreas} SRS simulation: Bias of fuzzy indicators at area
level. Areas sorted by increasing sample size}

\end{figure}

\begin{figure}[H]
    \centering
    \includegraphics[width = .6\textwidth, height = 0.3\textheight]{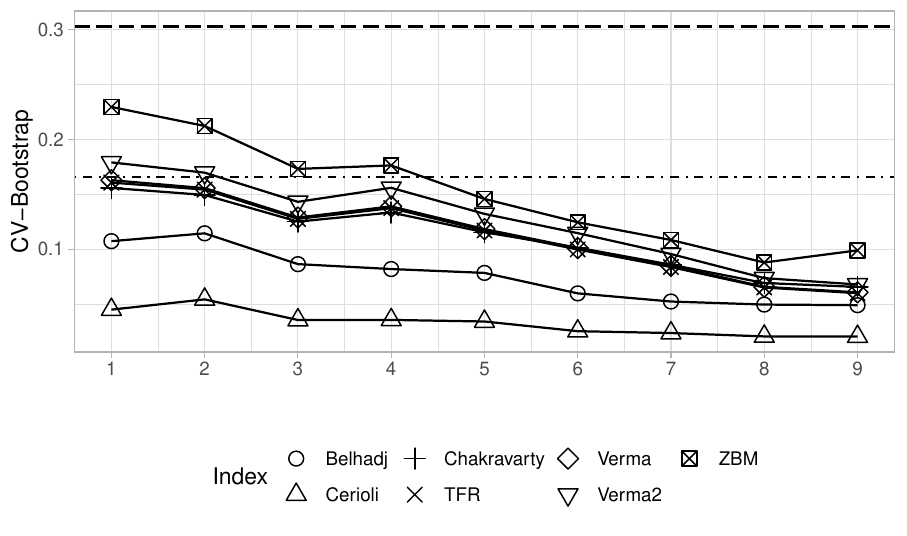}
    \caption{SRS simulation: Coefficient of Variation of fuzzy indices at area
level. Areas sorted by increasing sample size}
    \label{fig-CvSimpleAreas}
\end{figure}

As is possible to appreciate the bias is also very low in every areas, i.e. with every sample
size, and for each indices. Figure ~\ref{fig-BiasSimpleAreas} shows
 that even for the smaller domains the estimators are approximately
unbiased. The major insights that emerge from the picture is that using
Equation~\ref{eq-cerioli} the bias, although extremely contained is
always negative, while for three distributional indices the
figures are almost identical. Moving to the reliability of the estimates we note that CV (Figure \ref{fig-CvSimpleAreas}) and CV2 decrease as the
sample size increase. The CV2 which, in our view, is more interpretable
tends to be quite high $(>0.16)$ only when the sample size is
lower than 50. In particular, the \citep{cerioli1990fuzzyart} index is the ones with the
lowest CV2, always $(<0.06)$ while \citep{zedini2015new} is that with the
highest value. The distribution based indices
performs almost identically both in terms of bias, CV and CV2. In
the end, the Chackravarty (2019) index although based on an axiomatic approach, it attain slightly worse results in terms of CV and CV2.

\begin{table}[H]
\centering
\footnotesize
\begin{tabular}{p{0.5cm}p{1.5cm}p{1.5cm}p{1.5cm}p{1.5cm}p{1.5cm}p{1.5cm}p{1.5cm}} 
\hline
$n_i$ & Belhadj (2014)  & Cerioli and Zani (1999) & Chacravarty (2019) & Cheli and Lemmi (1995) & Betti and Verma (1999) & Betti et. al (2006) & Zedini and Belhadj (2015)  \\  
\hline
\multicolumn{8}{c}{Bias}\\
  \hline
  33 & 0.0014 & -0.0013 & 0.0005 & -0.0002 & -0.0002 & -0.0006 & -0.0685 \\ 
  42 & 0.0027 & -0.0021 & 0.0000 & -0.0012 & -0.0012 & -0.0018 & -0.2522 \\ 
  43 & -0.0000 & -0.0008 & 0.0001 & -0.0003 & -0.0003 & -0.0005 & -0.1354 \\ 
  65 & 0.0014 & -0.0016 & 0.0007 & 0.0006 & 0.0006 & 0.0008 & -0.0347 \\ 
  77 & 0.0009 & -0.0027 & 0.0017 & 0.0015 & 0.0015 & 0.0018 & -0.0735 \\ 
  110 & 0.0005 & -0.0020 & 0.0012 & 0.0011 & 0.0011 & 0.0012 & 0.0067 \\ 
  113 & -0.0009 & -0.0024 & -0.0006 & -0.0009 & -0.0009 & -0.0009 & -0.0046 \\ 
  167& 0.0012 & -0.0024 & -0.0001 & -0.0003 & -0.0003 & -0.0003 & -0.0548 \\ 
  176 & 0.0011 & -0.0026 & -0.0001 & -0.0003 & -0.0003 & -0.0002 & 0.0079 \\ 
 821  & 0.0015  & $<0.0001$ & $<0.0001$ & $<0.0001$ & $<0.0001$ & $<0.0001$ & $<0.0001$  \\ 
  \hline
\multicolumn{8}{c}{CV}\\
  \hline
33 & 0.1074 & 0.0450 & 0.1559 & 0.1607 & 0.1629 & 0.1791 & 0.4078 \\ 
  42 & 0.1145 & 0.0543 & 0.1494 & 0.1543 & 0.1557 & 0.1698 & 0.3243 \\ 
  43& 0.0864 & 0.0355 & 0.1252 & 0.1277 & 0.1289 & 0.1432 & 0.3154 \\ 
  65 & 0.0820 & 0.0356 & 0.1335 & 0.1374 & 0.1391 & 0.1560 & 0.3191 \\ 
  77 & 0.0785 & 0.0342 & 0.1155 & 0.1170 & 0.1184 & 0.1322 & 0.2589 \\ 
  110 & 0.0598 & 0.0254 & 0.1014 & 0.0998 & 0.1011 & 0.1147 & 0.2621 \\ 
113 & 0.0524 & 0.0237 & 0.0859 & 0.0838 & 0.0849 & 0.0957 & 0.2144 \\ 
  167 & 0.0496 & 0.0206 & 0.0692 & 0.0651 & 0.0658 & 0.0737 & 0.1500 \\ 
  176 & 0.0491 & 0.0204 & 0.0657 & 0.0600 & 0.0608 & 0.0679 & 0.1478 \\
  821 & 0.0220  & 0.0094& 0.0323 & 0.0032 & 0.0034  &  0.0155&  0.0164\\
    \hline
 \multicolumn{8}{c}{CV2}\\
   \hline
33 & 0.1068 & 0.0449 & 0.1540 & 0.1587 & 0.1608 & 0.1763 & 0.3776 \\ 
42 & 0.1138 & 0.0543 & 0.1478 & 0.1525 & 0.1538 & 0.1674 & 0.3085 \\ 
  43 & 0.0861 & 0.0355 & 0.1242 & 0.1266 & 0.1278 & 0.1417 & 0.3008 \\ 
  65& 0.0817 & 0.0356 & 0.1323 & 0.1361 & 0.1377 & 0.1541 & 0.3040 \\ 
  77 & 0.0782 & 0.0342 & 0.1147 & 0.1162 & 0.1176 & 0.1311 & 0.2507 \\ 
  110 & 0.0597 & 0.0253 & 0.1009 & 0.0993 & 0.1005 & 0.1139 & 0.2535 \\ 
  113 & 0.0523 & 0.0237 & 0.0856 & 0.0835 & 0.0846 & 0.0952 & 0.2096 \\ 
  167 & 0.0495 & 0.0206 & 0.0691 & 0.0650 & 0.0656 & 0.0735 & 0.1483 \\ 
  176 & 0.0490 & 0.0204 & 0.0655 & 0.0599 & 0.0606 & 0.0678 & 0.1463 \\ 
 821 & 0.0219 & 0.0094 & 0.0323 & 0.0032 & 0.0034 & 0.0155  & 0.0163\\
   \hline
\end{tabular}
\caption{\label{tab1}SRS simulation: Bias, Coefficient of Variation, Coefficient of Variation of second order}
\end{table}


\begin{figure}[H]

\begin{minipage}[t]{.3\textwidth}

{\centering 

\raisebox{-\height}{

\includegraphics[width=\textwidth]{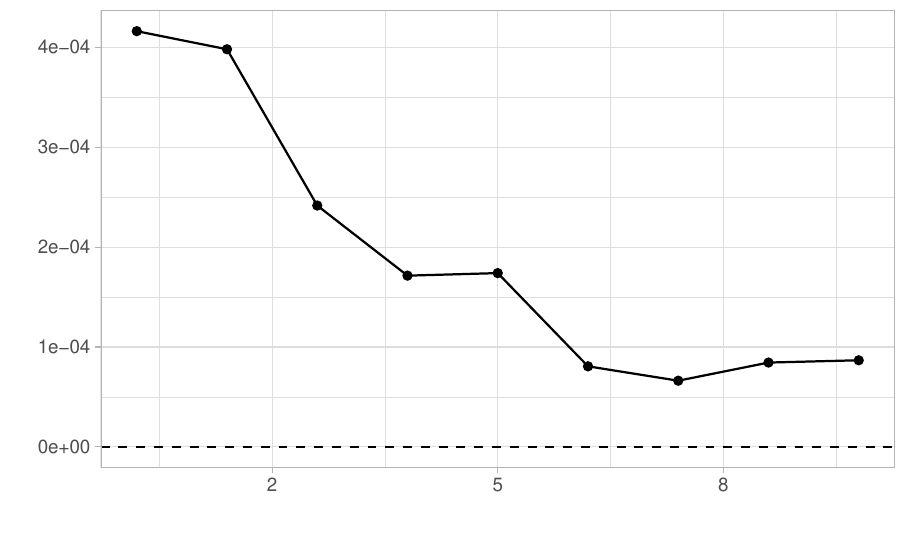}

}

}

\subcaption{\label{fig-MseSimpleAreas-1}Belhadj (2014)}
\end{minipage}%
\begin{minipage}[t]{.3\textwidth}

{\centering 

\raisebox{-\height}{

\includegraphics[width=\textwidth]{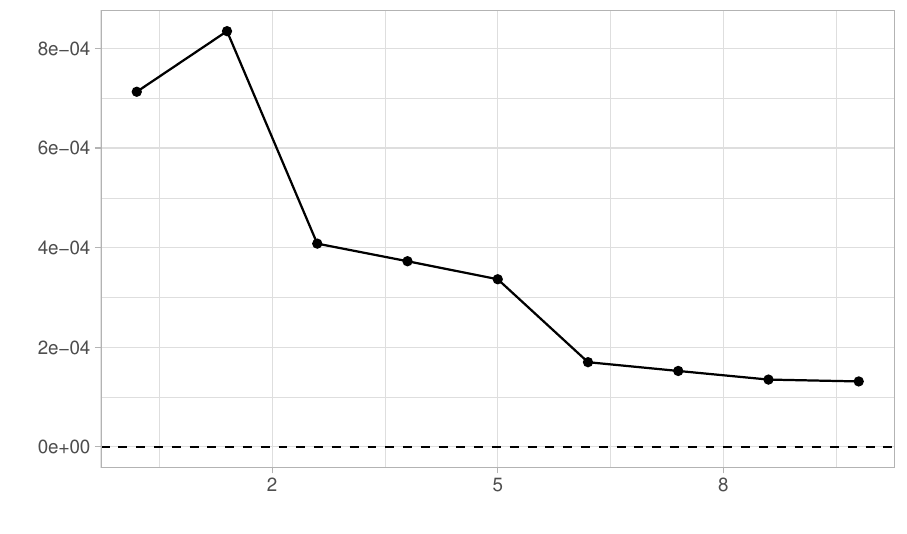}

}

}

\subcaption{\label{fig-MseSimpleAreas-2}Cerioli and Zani (1999)}
\end{minipage}%
\begin{minipage}[t]{.3\textwidth}

{\centering 

\raisebox{-\height}{

\includegraphics[width=\textwidth]{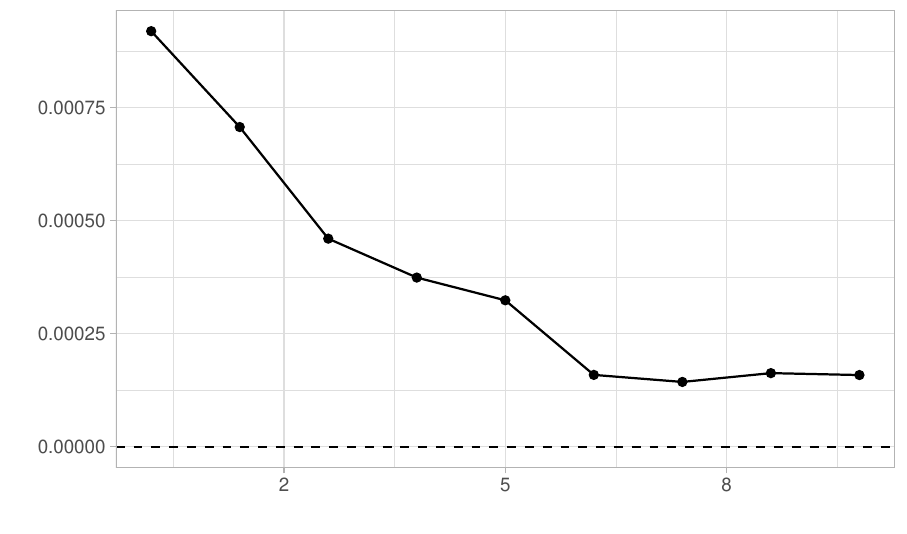}

}

}

\subcaption{\label{fig-MseSimpleAreas-3}Chackravarty (2019)}
\end{minipage}%
\newline
\begin{minipage}[t]{.3\textwidth}

{\centering 

\raisebox{-\height}{

\includegraphics[width=\textwidth]{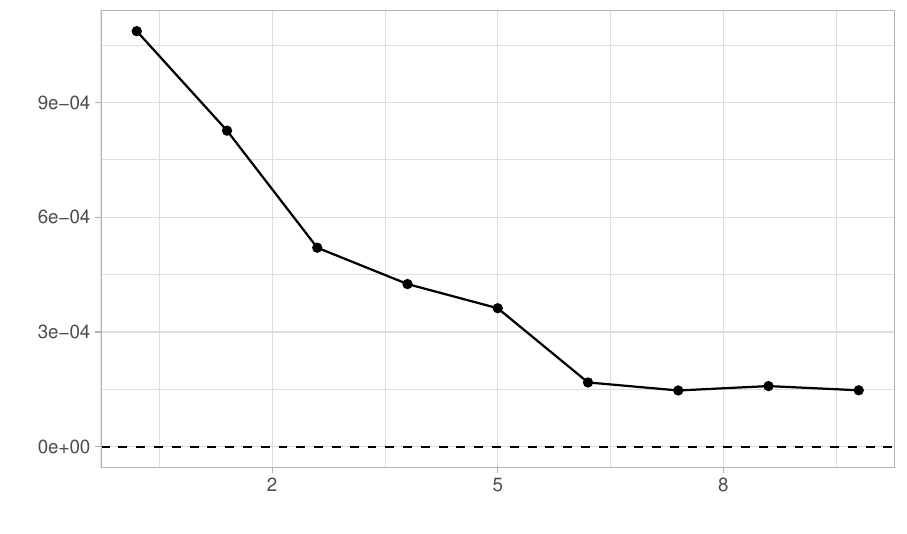}

}

}

\subcaption{\label{fig-MseSimpleAreas-4}Cheli and Lemmi (1995)}
\end{minipage}%
\begin{minipage}[t]{.3\textwidth}

{\centering 

\raisebox{-\height}{

\includegraphics[width=\textwidth]{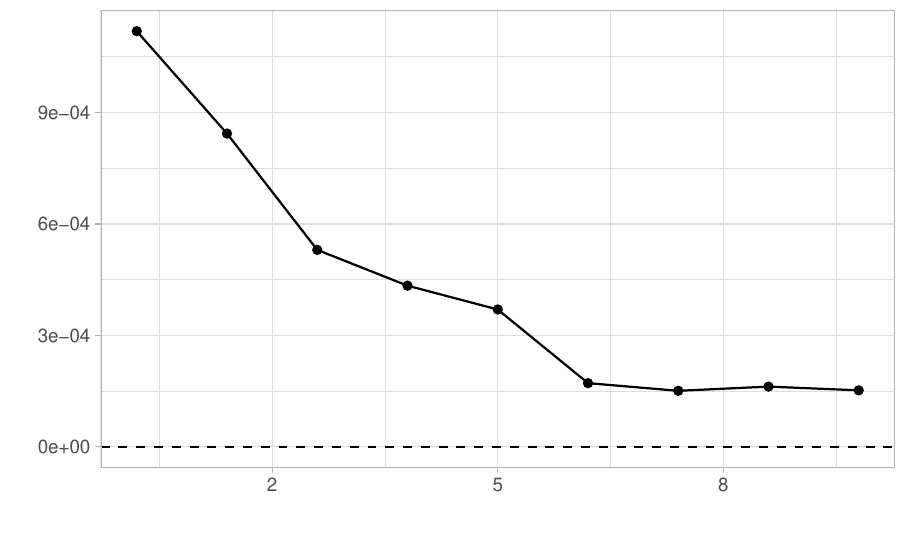}

}

}

\subcaption{\label{fig-MseSimpleAreas-5}Betti and Verma (1999)}
\end{minipage}%
\begin{minipage}[t]{.3\textwidth}

{\centering 

\raisebox{-\height}{

\includegraphics[width=\textwidth]{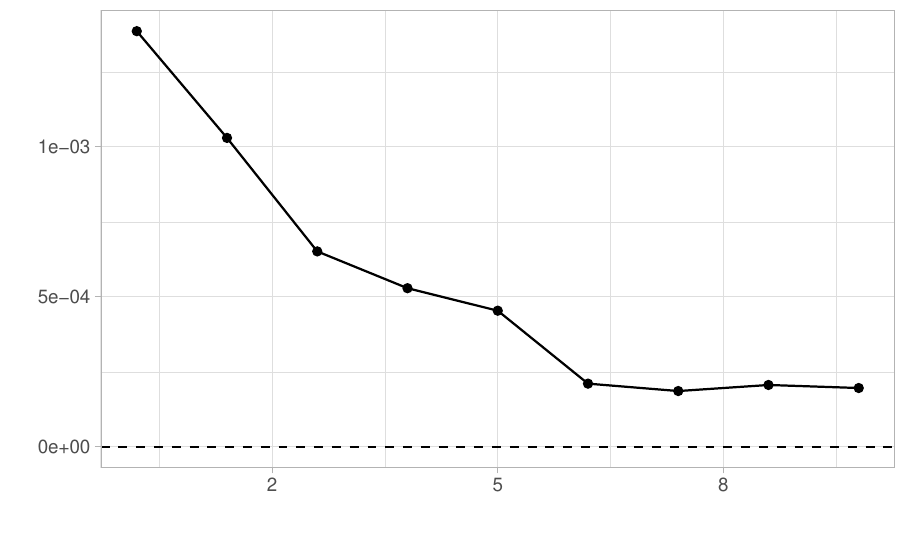}

}

}

\subcaption{\label{fig-MseSimpleAreas-6}Betti et. al (2006)}
\end{minipage}
\newline

\begin{minipage}[t]{.3\textwidth}

{\centering 

\raisebox{-\height}{

\includegraphics[width=\textwidth]{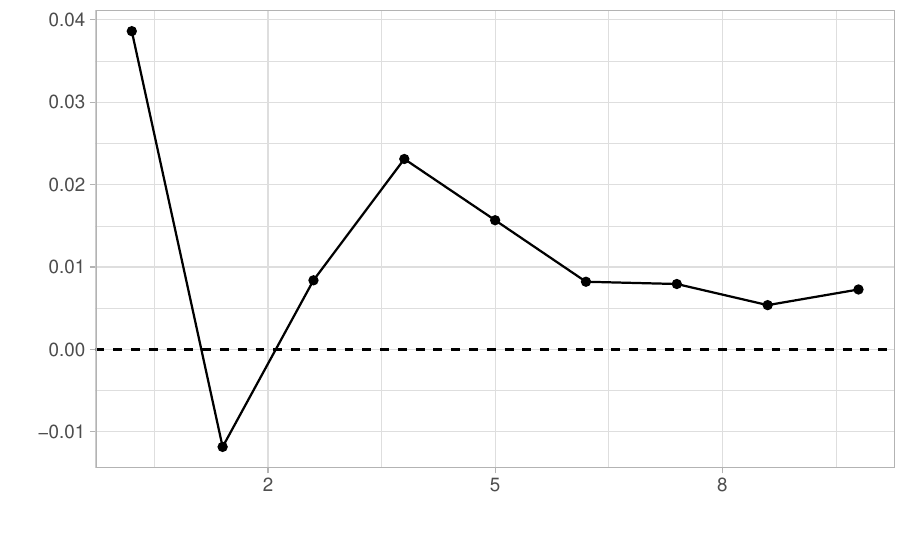}

}

}

\subcaption{\label{fig-MseSimpleAreasZBM}Zedini and Belhadj (2015)}
\end{minipage}%

\caption{\label{fig-MseSimpleAreas}SRS simulation: Bias of MSE estimator for the fuzzy indicators at area
level. Areas sorted by increasing sample size}

\end{figure}

\noindent
Moving to the MSE estimation, Table \ref{tab2} reports the values of the
ATMSE, AEMSE and the ABMSE. 
Values for each areas of the ABMSE are reported in
Figure~\ref{fig-MseSimpleAreas}. Not surprisingly the non-parametric
bootstrap estimates have an ABMSE that approaches zero as the sample
size increases. In general the values of the ATMSE is very similar to
the ones of the ABMSE for every estimators. There are not remarkable
difference in the performance of the bootstrap estimator between one
index and another. We also notice that for those areas that have low
sample size $(<50)$ the variance is overestimated for all the index except that for one single area for the index defined by Zedini and Belhadj (2015). This is not
necessarily a problem as the overestimation of variance can be
considered as conservative. In the next paragraph, with the complex
design simulation, we will verify what happens if the sample size is
even lower and, in case of higher variability we will suggest ad-hoc
methods to estimates those quantities in a reliable way.

\begin{table}[H]
\centering
\footnotesize
\begin{tabular}{p{0.5cm}p{1.5cm}p{1.5cm}p{1.5cm}p{1.5cm}p{1.5cm}p{1.5cm}p{1.5cm}}  
\hline
$n_i$ & Belhadj (2014)  & Cerioli and Zani (1999) & Chacravarty (2019) & Cheli and Lemmi (1995) & Betti and Verma (1999) & Betti et. al (2006) & Zedini and Belhadj (2015)  \\    \hline

\multicolumn{8}{c}{ATMSE}\\
  \hline
 
33 & 2.039E-06 & 1.621E-06 & 2.307E-07 & 3.956E-08 & 5.009E-08 & 3.385E-07 & 4.695E-03 \\ 
  42 & 7.089E-06 & 4.244E-06 & 2.489E-10 & 1.442E-06 & 1.557E-06 & 3.326E-06 & 6.361E-02 \\ 
  43 & 1.326E-09 & 5.782E-07 & 4.584E-09 & 1.051E-07 & 1.157E-07 & 2.676E-07 & 1.833E-02 \\ 
  65 & 1.976E-06 & 2.594E-06 & 5.131E-07 & 3.744E-07 & 3.920E-07 & 5.846E-07 & 1.204E-03 \\ 
  77 & 7.690E-07 & 7.426E-06 & 2.869E-06 & 2.104E-06 & 2.220E-06 & 3.221E-06 & 5.403E-03 \\ 
  110 & 2.230E-07 & 3.896E-06 & 1.389E-06 & 1.171E-06 & 1.188E-06 & 1.467E-06 & 4.502E-05 \\ 
  113 & 7.399E-07 & 5.966E-06 & 3.073E-07 & 7.268E-07 & 7.374E-07 & 7.462E-07 & 2.101E-05 \\ 
  167 & 1.466E-06 & 5.704E-06 & 3.239E-09 & 1.057E-07 & 1.062E-07 & 1.059E-07 & 3.007E-03 \\ 
  176 & 1.147E-06 & 6.649E-06 & 1.378E-08 & 1.222E-07 & 1.141E-07 & 5.382E-08 & 6.295E-05 \\  
  821 & 2.186E-06 & 8.283E-08 & 4.468E-09 &  1.945E-07 &1.943E-07 & 1.945E-07 &  1.914E-06\\ 

   \hline
\multicolumn{8}{c}{AEMSE}\\
  \hline
33 & 4.181E-04 & 7.146E-04 & 9.198E-04 & 1.086E-03 & 1.119E-03 & 1.386E-03 & 1.366E-01 \\ 
  42 & 4.050E-04 & 8.389E-04 & 7.075E-04 & 8.273E-04 & 8.448E-04 & 1.033E-03 & 1.208E-01 \\ 
  43 & 2.416E-04 & 4.087E-04 & 4.606E-04 & 5.203E-04 & 5.303E-04 & 6.518E-04 & 8.858E-02 \\ 
  65 & 1.734E-04 & 3.753E-04 & 3.750E-04 & 4.256E-04 & 4.343E-04 & 5.294E-04 & 7.960E-02 \\ 
  77 & 1.747E-04 & 3.440E-04 & 3.271E-04 & 3.642E-04 & 3.720E-04 & 4.571E-04 & 6.644E-02 \\ 
  110 & 8.086E-05 & 1.740E-04 & 1.607E-04 & 1.695E-04 & 1.728E-04 & 2.122E-04 & 3.658E-02 \\ 
  113 & 6.699E-05 & 1.584E-04 & 1.440E-04 & 1.481E-04 & 1.517E-04 & 1.869E-04 & 3.111E-02 \\ 
  167 & 8.587E-05 & 1.409E-04 & 1.632E-04 & 1.589E-04 & 1.624E-04 & 2.065E-04 & 2.439E-02 \\ 
  176 & 8.784E-05 & 1.381E-04 & 1.590E-04 & 1.480E-04 & 1.524E-04 & 1.963E-04 & 1.637E-02 \\ 
  821 & 1.460E-05 & 2.740E-05 & 2.811E-05 & 2.948E-07 & 3.442E-07 & 7.102E-06 & 4.102E-04 \\ 
    \hline
 \multicolumn{8}{c}{ABMSE}\\
   \hline
33 & 0.0004 & 0.0007 & 0.0009 & 0.0011 & 0.0011 & 0.0014 & 0.1319 \\ 
  42 & 0.0004 & 0.0008 & 0.0007 & 0.0008 & 0.0008 & 0.0010 & 0.0572 \\ 
  43 & 0.0002 & 0.0004 & 0.0005 & 0.0005 & 0.0005 & 0.0007 & 0.0702 \\ 
  65 & 0.0002 & 0.0004 & 0.0004 & 0.0004 & 0.0004 & 0.0005 & 0.0784 \\ 
  77 & 0.0002 & 0.0003 & 0.0003 & 0.0004 & 0.0004 & 0.0005 & 0.0610 \\ 
  110 & 0.0001 & 0.0002 & 0.0002 & 0.0002 & 0.0002 & 0.0002 & 0.0365 \\ 
  113 & 0.0001 & 0.0002 & 0.0001 & 0.0001 & 0.0002 & 0.0002 & 0.0311 \\ 
  167 & 0.0001 & 0.0001 & 0.0002 & 0.0002 & 0.0002 & 0.0002 & 0.0214 \\ 
  176 & 0.0001 & 0.0001 & 0.0002 & 0.0001 & 0.0002 & 0.0002 & 0.0163 \\ 
  821& 1.241E-05 & 2.732E-05 & 2.810E-05 & 1.001E-07 & 1.498E-07 & 6.900E-06  & 0.0013\\ 
   \hline
\end{tabular}
\caption{\label{tab2}SRS simulation: ATMSE, AEMSE and ABMSE.}
\end{table}

The mean squared error is contained for each method considered with
values that are very close to zero. However, what is more important in
sample surveys is the coefficient of variation of the estimates
obtained. In fact, estimates from sample survey are usually considered
reliable depending on how much is low/high coefficient of variation. 
\citep{Canada} set as upper limit for the publication a CV equal to 0.166 which we draw in the Figure~\ref{fig-CvSimpleAreas}. In this case, we see that the indices
that use the membership functions Equation~\ref{eq-belhadj2015} and
Equation~\ref{eq-chackra} are those that have higher CV but still can be
estimated with high accuracy.

\noindent
Regarding the estimation of the mean squared errors
Figure~\ref{fig-MseSimpleAreas} shows that - not surprisingly - this
approaches zero as the sample size increases. We also notice that for
those areas that have low sample size, i.e. $<50$, the variance is
overestimated. This is not necessarily a problem as the overestimation
of variance can be considered as conservative.

Figure~\ref{fig-CvSimpleAreas} compares the coefficient of variation of
each method and suggest that the method by \citep{cerioli1990fuzzyart} is
the one that is estimated with higher accuracy (less than 20\%). Other
methods tend to be estimated with higher CVs especially for small areas.
This means that from an official statistics perspective these estimates
would not be publishable. However, we should keep in mind that the
results above show that these high CVs may be an outcome of an
over-estimation of the variance.

\hypertarget{complex-design}{%
\subsection{Complex-design}\label{complex-design}}

We repeat the same analysis show above assuming that the survey is
carried out via a complex sampling design. We report results either for
the bootstrap and jackknife estimators of the variance. The main point
is to understand if the knowledge of the sample scheme, mandatory for
the jackknife estimator, is necessary to obtain reliable estimates or
if, is enough, to perform a faster and easier non-parametric bootstrap.
As previously done we will start discussing the bias of the estimates,
after which we will concentrate on the MSE estimation.

\begin{figure}[H]

\begin{minipage}[t]{.3\textwidth}

{\centering 

\raisebox{-\height}{

\includegraphics[width=\textwidth]{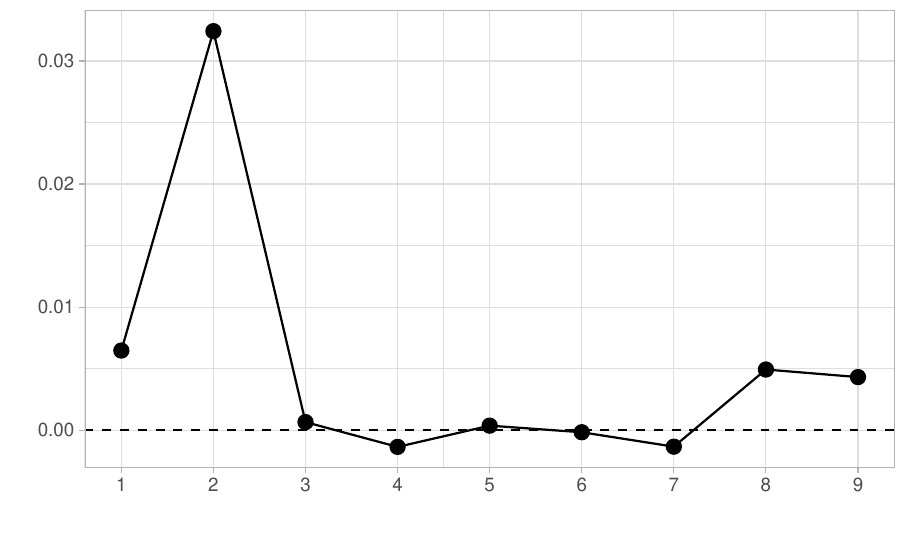}

}

}

\subcaption{\label{fig-BiasSimpleAreasCompl-1}Belhadj (2014)}
\end{minipage}%
\begin{minipage}[t]{.3\textwidth}

{\centering 

\raisebox{-\height}{

\includegraphics[width=\textwidth]{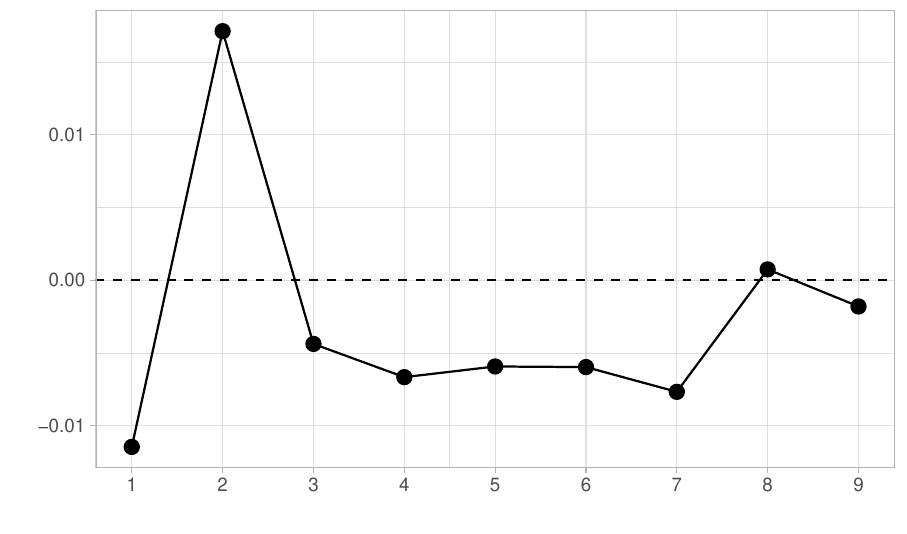}

}

}

\subcaption{\label{fig-BiasSimpleAreasCompl-2}Cerioli and Zani (1999)}
\end{minipage}%
\begin{minipage}[t]{.3\textwidth}

{\centering 

\raisebox{-\height}{

\includegraphics[width=\textwidth]{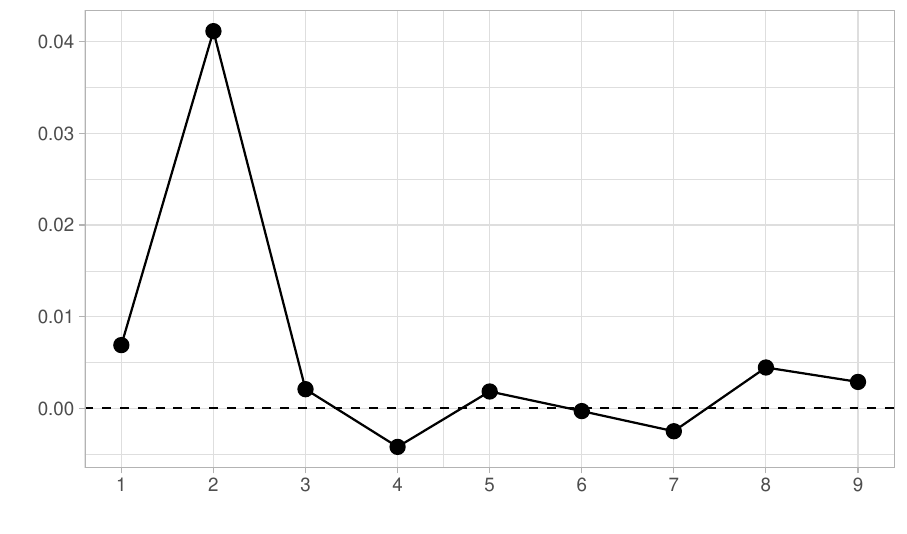}

}

}

\subcaption{\label{fig-BiasSimpleAreasCompl-3}Chackravarty (2019)}
\end{minipage}%
\newline
\begin{minipage}[t]{.3\textwidth}

{\centering 

\raisebox{-\height}{

\includegraphics[width=\textwidth]{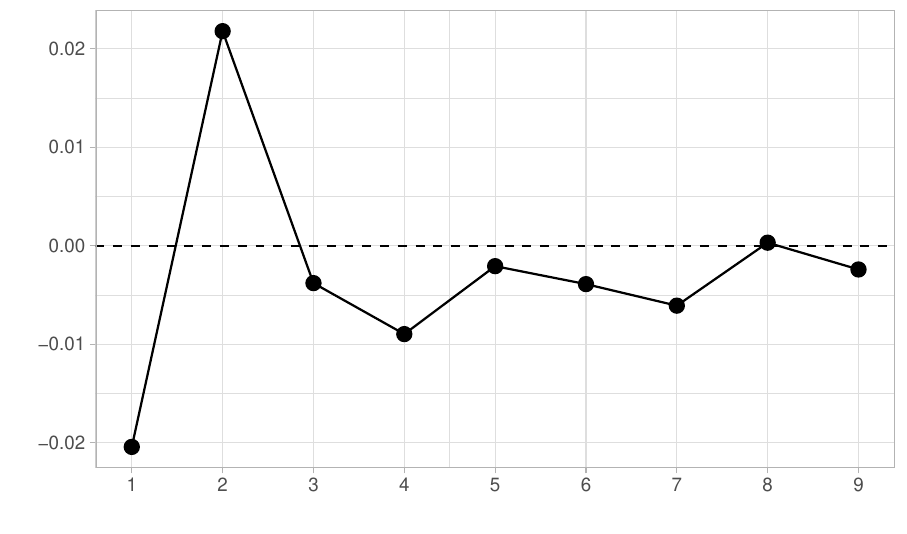}

}

}

\subcaption{\label{fig-BiasSimpleAreasCompl-4}Cheli and Lemmi (1995)}
\end{minipage}%
\begin{minipage}[t]{.3\textwidth}

{\centering 

\raisebox{-\height}{

\includegraphics[width=\textwidth]{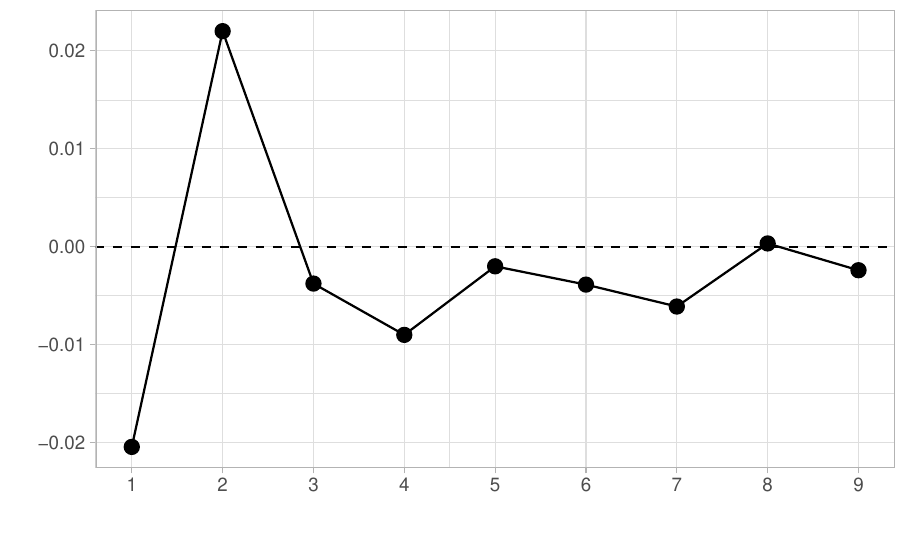}

}

}

\subcaption{\label{fig-BiasSimpleAreasCompl-5}Betti and Verma (1999)}
\end{minipage}%
\begin{minipage}[t]{.3\textwidth}

{\centering 

\raisebox{-\height}{

\includegraphics[width=\textwidth]{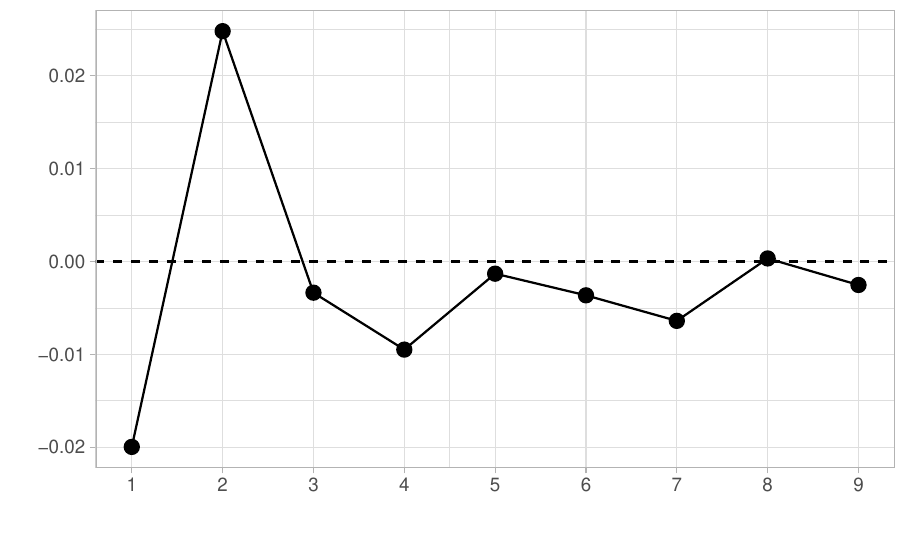}

}

}

\subcaption{\label{fig-BiasSimpleAreasCompl-6}Betti et. al, (2006)}
\end{minipage}
\newline

\begin{minipage}[t]{.3\textwidth}

{\centering 

\raisebox{-\height}{

\includegraphics[width=\textwidth]{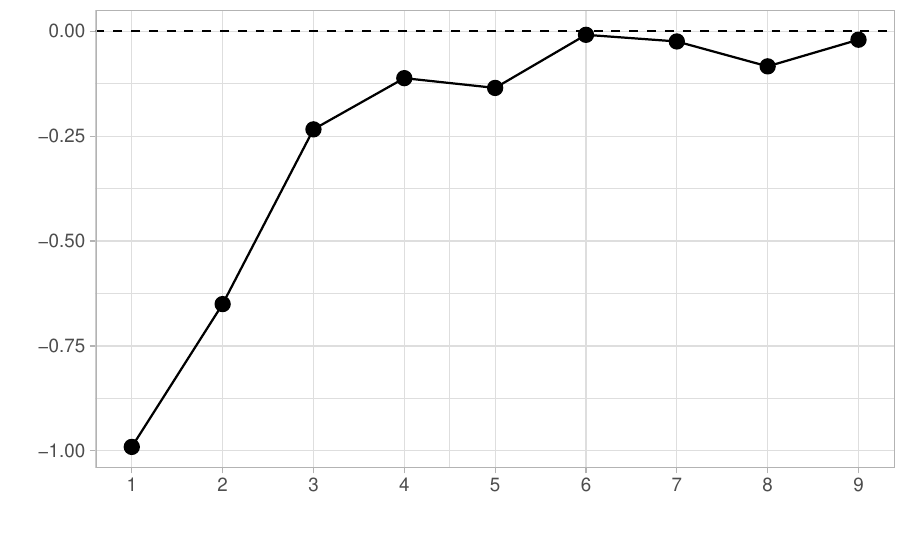}

}

}

\subcaption{\label{fig-BiasSimpleAreasComplZBM}Zedini and Belhadj (2015)}
\end{minipage}%

\caption{\label{fig-BiasSimpleAreasCompl}Complex simulation: Bias of fuzzy indicators at area
level. Areas sorted by increasing sample size}

\end{figure}

Table \ref{tab3} reports the value of Bias, CV and CV2. Again, also
in this scenario, the bias is very low for every index and clearly
tends to 0 if the sample size increase
(Figure~\ref{fig-BiasSimpleAreasCompl}). Two indices deserve for more attention. The Cerioli and Zani (1999) it is not always negative biased as in the previous simulation. Zedini and Belhadj (2015) is the index  most affected by the sample size. This index tends to be unbiased with quite high sample size, i.e. $>50$, but can be severally biased for small sample size.

\begin{figure}[H]

\begin{minipage}[t]{.5\textwidth}

{\centering 

\raisebox{-\height}{

\includegraphics[width=\textwidth]{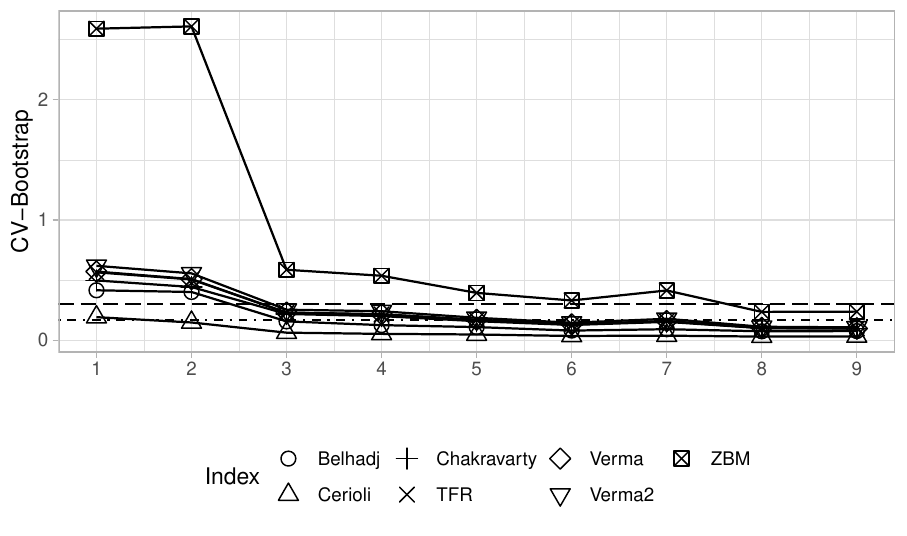}

}

}

\subcaption{\label{fig-CvSimpleAreasCompl-1}Bootstrap}
\end{minipage}%
\begin{minipage}[t]{.5\textwidth}

{\centering 

\raisebox{-\height}{

\includegraphics[width=\textwidth]{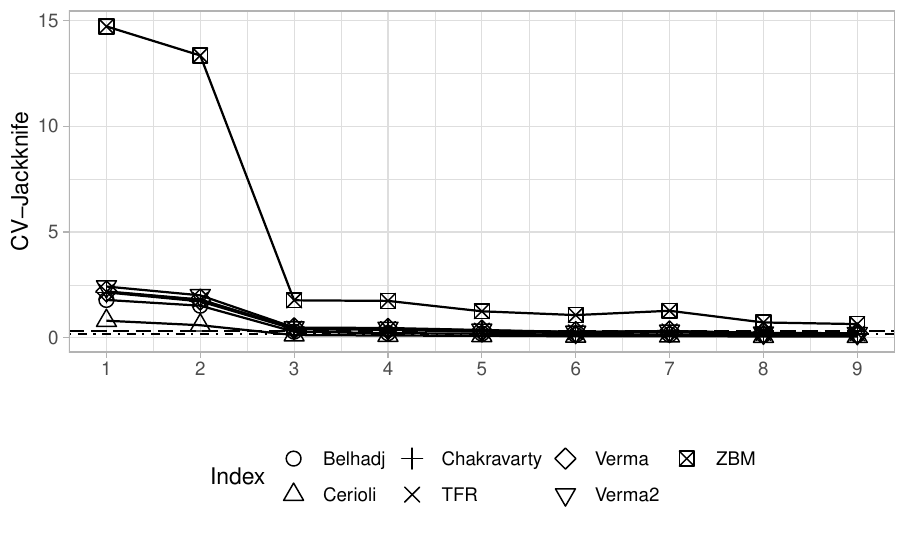}

}

}

\subcaption{\label{fig-CvSimpleAreasCompl-2}Jackknife}
\end{minipage}%

\caption{\label{fig-CvSimpleAreasCompl}Complex simulation: Coefficient of Variation for each  index at area
level. Areas sorted by increasing sample size}

\end{figure}

In this case, having to compare
two different estimators for the MSE we have reported in
Figure~\ref{fig-CvSimpleAreasCompl} both the CV for the bootstrap and
for jackknife. Speaking, for the moment, only about the variability we note that
the two MSE estimators are almost identical in the trends,
as confirmed also by the CV2 reported in the table, but the jackknife
estimates are greater than the ones obtained with the bootstrap. From
Figure~\ref{fig-CvSimpleAreasCompl} it is, however, important to come to
other two conclusions. First of all, the CV (CV2) becomes lower if the
sample size increase for every estimator and that the Cerioli and Zani (1999) estimator
is the one with the lowest CV. Again, as previously said, the
Chackravarty (2019)  performs very similar to the distribution based indices and worst than the Belhadj (2014) one.

\begin{table}[H]
\centering
\footnotesize
\begin{tabular}{p{0.5cm}p{1.5cm}p{1.5cm}p{1.5cm}p{1.5cm}p{1.5cm}p{1.5cm}p{1.5cm}}  
\hline
$n_i$ & Belhadj (2014)  & Cerioli and Zani (1999) & Chacravarty (2019) & Cheli and Lemmi (1995) & Betti and Verma (1999) & Betti et. al (2006) & Zedini and Belhadj (2015)  \\  
  \hline
\multicolumn{8}{c}{Bias}\\
  \hline

3 & 0.0065 & -0.0115 & 0.0069 & -0.0204 & -0.0205 & -0.0200 & -0.9905 \\ 
 6 & 0.0324 & 0.0171 & 0.0411 & 0.0218 & 0.0220 & 0.0248 & -0.6502 \\ 
  48 & 0.0007 & -0.0044 & 0.0021 & -0.0038 & -0.0038 & -0.0034 & -0.2339 \\ 
  55& -0.0014 & -0.0067 & -0.0042 & -0.0090 & -0.0090 & -0.0095 & -0.1120 \\ 
  100 & 0.0004 & -0.0059 & 0.0018 & -0.0021 & -0.0020 & -0.0013 & -0.1352 \\ 
  124 & -0.0002 & -0.0060 & -0.0003 & -0.0039 & -0.0039 & -0.0037 & -0.0087 \\ 
  128& -0.0013 & -0.0077 & -0.0025 & -0.0061 & -0.0061 & -0.0064 & -0.0245 \\ 
  157 & 0.0049 & 0.0007 & 0.0045 & 0.0003 & 0.0003 & 0.0003 & -0.0839 \\ 
  187 & 0.0043 & -0.0018 & 0.0029 & -0.0024 & -0.0024 & -0.0025 & -0.0203 \\ 
  808 & 0.0028 & -0.0030 & -0.0013 &-0.0032  &  -0.0033&  -0.0032& -0.0016\\ 
  \hline
\multicolumn{8}{c}{CV-Bootstrap}\\
  \hline
3 & 0.4154 & 0.1921 & 0.4971 & 0.5649 & 0.5711 & 0.6189 & 2.5906 \\ 
  6 & 0.4014 & 0.1478 & 0.4427 & 0.5054 & 0.5085 & 0.5560 & 2.6096 \\ 
  48 & 0.1568 & 0.0630 & 0.2184 & 0.2289 & 0.2303 & 0.2540 & 0.5860 \\ 
  55& 0.1257 & 0.0529 & 0.2008 & 0.2147 & 0.2166 & 0.2412 & 0.5362 \\ 
  100 & 0.1094 & 0.0477 & 0.1584 & 0.1685 & 0.1701 & 0.1867 & 0.3937 \\ 
  124 & 0.0804 & 0.0355 & 0.1269 & 0.1314 & 0.1325 & 0.1459 & 0.3306 \\ 
  128 & 0.0917 & 0.0381 & 0.1527 & 0.1581 & 0.1598 & 0.1784 & 0.4139 \\ 
  157 & 0.0745 & 0.0303 & 0.1005 & 0.1018 & 0.1027 & 0.1120 & 0.2357 \\ 
  187 & 0.0762 & 0.0313 & 0.0987 & 0.0975 & 0.0983 & 0.1071 & 0.2379 \\ 
  808 & 0.0420 & 0.0175 & 0.0595 & 0.0136 & 0.0132 & 0.0282 &  0.0569\\ 
    \hline
\multicolumn{8}{c}{CV-Jackknife}\\
  \hline
3 & 1.7850 & 0.8024 & 2.1429 & 2.1684 & 2.1909 & 2.4199 & 14.7225 \\ 
 6 & 1.5131 & 0.5889 & 1.7172 & 1.7858 & 1.8055 & 2.0024 & 13.3466 \\ 
48 & 0.2833 & 0.1232 & 0.4057 & 0.4372 & 0.4409 & 0.4791 & 1.7672 \\ 
  55 & 0.2142 & 0.1015 & 0.3704 & 0.4138 & 0.4174 & 0.4569 & 1.7402 \\ 
  100& 0.1911 & 0.0899 & 0.2927 & 0.3355 & 0.3376 & 0.3638 & 1.2483 \\ 
  124 & 0.1302 & 0.0660 & 0.2135 & 0.2595 & 0.2598 & 0.2692 & 1.0694 \\ 
  128 & 0.1475 & 0.0718 & 0.2563 & 0.2967 & 0.2981 & 0.3169 & 1.2728 \\ 
  157 & 0.1285 & 0.0552 & 0.1815 & 0.2173 & 0.2174 & 0.2268 & 0.7200 \\ 
  187 & 0.1309 & 0.0517 & 0.1680 & 0.2043 & 0.2041 & 0.2100 & 0.6439 \\ 
  808 & 0.1151 &  0.0473 & 0.1592 & 0.0232 & 0.0212 & 0.0740 & 0.0892\\
  \hline
 \multicolumn{8}{c}{CV2-Bootstrap}\\
   \hline
3 & 0.3836 & 0.1886 & 0.4452 & 0.4918 & 0.4959 & 0.5262 & 0.9329 \\ 
 6 & 0.3725 & 0.1462 & 0.4048 & 0.4510 & 0.4533 & 0.4860 & 0.9338 \\ 
48 & 0.1549 & 0.0629 & 0.2133 & 0.2231 & 0.2245 & 0.2462 & 0.5056 \\ 
  55 & 0.1247 & 0.0528 & 0.1968 & 0.2099 & 0.2117 & 0.2345 & 0.4726 \\ 
  100 & 0.1087 & 0.0476 & 0.1564 & 0.1662 & 0.1677 & 0.1836 & 0.3663 \\ 
  124 & 0.0801 & 0.0354 & 0.1259 & 0.1303 & 0.1314 & 0.1443 & 0.3139 \\ 
  128 & 0.0913 & 0.0380 & 0.1510 & 0.1561 & 0.1578 & 0.1756 & 0.3825 \\ 
  157 & 0.0743 & 0.0303 & 0.1000 & 0.1013 & 0.1022 & 0.1113 & 0.2294 \\ 
  187 & 0.0760 & 0.0313 & 0.0982 & 0.0970 & 0.0978 & 0.1065 & 0.2314 \\ 
  808 & 0.0419 & 0.0175 & 0.0593 & 0.0136 & 0.0132 & 0.0282 & 0.0543\\
   \hline
 \multicolumn{8}{c}{CV2-Jackknife}\\
   \hline
  3 & 0.8724 & 0.6258 & 0.9062 & 0.9081 & 0.9097 & 0.9242 & 0.9977 \\ 
  6 & 0.8343 & 0.5075 & 0.8641 & 0.8725 & 0.8748 & 0.8946 & 0.9972 \\ 
  48 & 0.2726 & 0.1223 & 0.3760 & 0.4006 & 0.4034 & 0.4321 & 0.8703 \\ 
  55 & 0.2094 & 0.1010 & 0.3474 & 0.3824 & 0.3852 & 0.4156 & 0.8670 \\ 
  100 & 0.1877 & 0.0896 & 0.2809 & 0.3181 & 0.3199 & 0.3419 & 0.7805 \\ 
  124 & 0.1291 & 0.0658 & 0.2088 & 0.2512 & 0.2515 & 0.2600 & 0.7304 \\ 
  128 & 0.1460 & 0.0716 & 0.2482 & 0.2845 & 0.2857 & 0.3021 & 0.7863 \\ 
  157 & 0.1275 & 0.0551 & 0.1785 & 0.2123 & 0.2125 & 0.2212 & 0.5843 \\ 
  187 & 0.1298 & 0.0516 & 0.1656 & 0.2002 & 0.2000 & 0.2055 & 0.5414 \\ 
 808 & 0.1149 &  0.0472 & 0.1573 & 0.0226 & 0.0212 & 0.0738 & 0.0792\\ 
 \hline
\end{tabular}
\caption{\label{tab3}Complex simulation: Bias, Coefficient of Variation and Coefficient of Variation of second order distinguished for bootstrap and jackknife estimates.}
\end{table}


\noindent

\begin{figure}[H]

\begin{minipage}[t]{.3\textwidth}

{\centering 

\raisebox{-\height}{

\includegraphics[width=\textwidth]{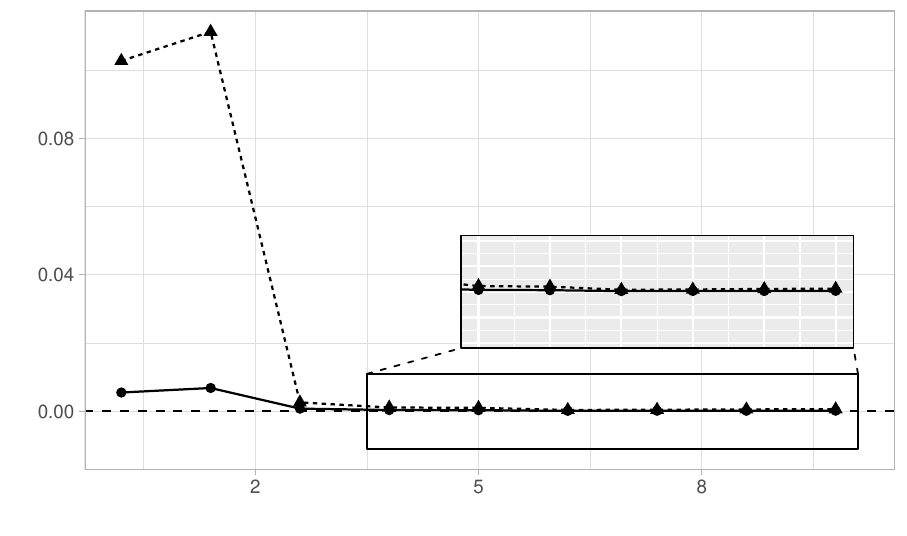}

}

}

\subcaption{\label{fig-MseSimpleAreasComp-1}Belhadj (2014)}
\end{minipage}%
\begin{minipage}[t]{.3\textwidth}

{\centering 

\raisebox{-\height}{

\includegraphics[width=\textwidth]{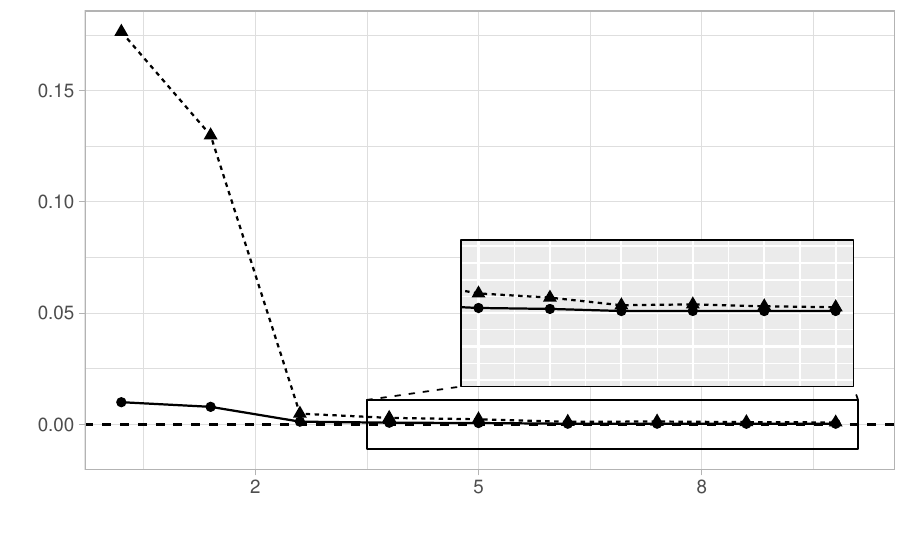}

}

}

\subcaption{\label{fig-MseSimpleAreasComp-2}Cerioli and Zani (1999)}
\end{minipage}%
\begin{minipage}[t]{.3\textwidth}

{\centering 

\raisebox{-\height}{

\includegraphics[width=\textwidth]{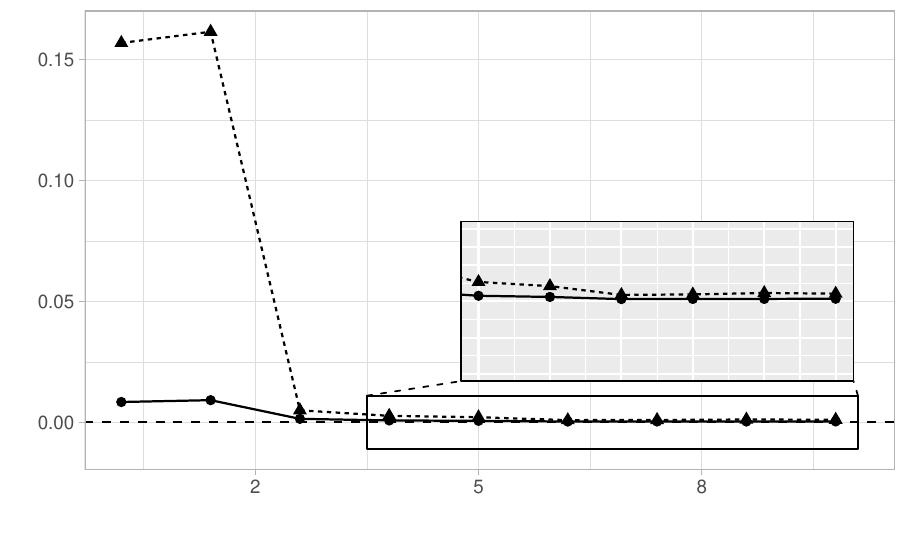}

}

}

\subcaption{\label{fig-MseSimpleAreasComp-3}Chackravarty (2019)}
\end{minipage}%
\newline
\begin{minipage}[t]{.3\textwidth}

{\centering 

\raisebox{-\height}{

\includegraphics[width=\textwidth]{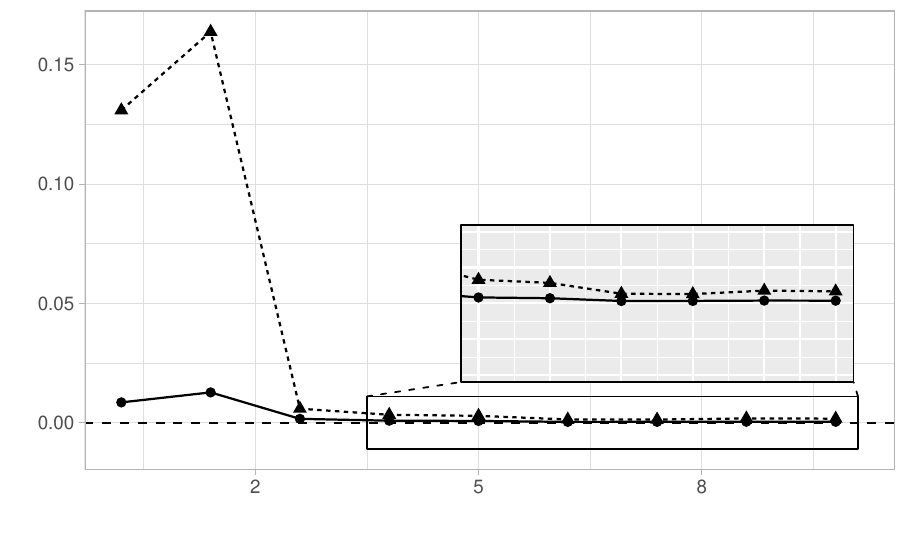}

}

}

\subcaption{\label{fig-MseSimpleAreasComp-4}Cheli and Lemmi (1995)}
\end{minipage}%
\begin{minipage}[t]{.3\textwidth}

{\centering 

\raisebox{-\height}{

\includegraphics[width=\textwidth]{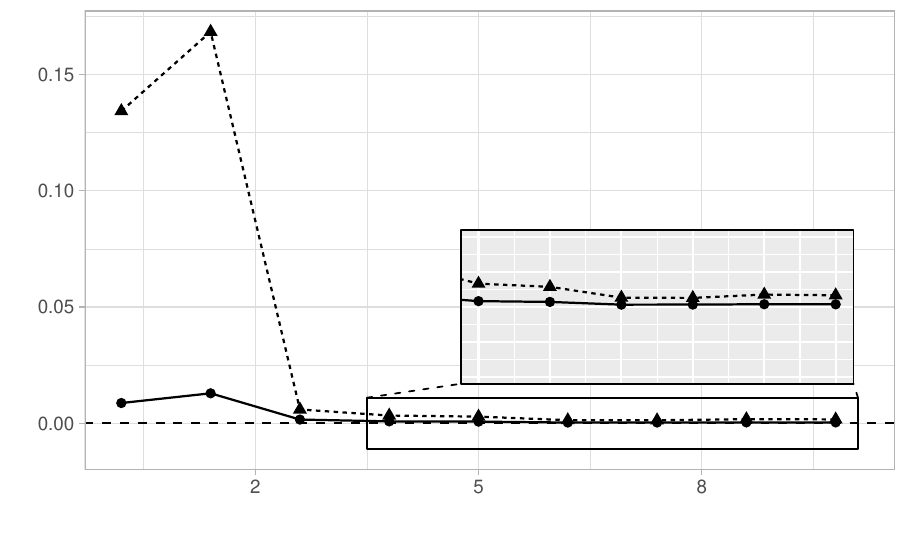}

}

}

\subcaption{\label{fig-MseSimpleAreasComp-5}Betti and Verma (1999)}
\end{minipage}%
\begin{minipage}[t]{.3\textwidth}

{\centering 

\raisebox{-\height}{

\includegraphics[width=\textwidth]{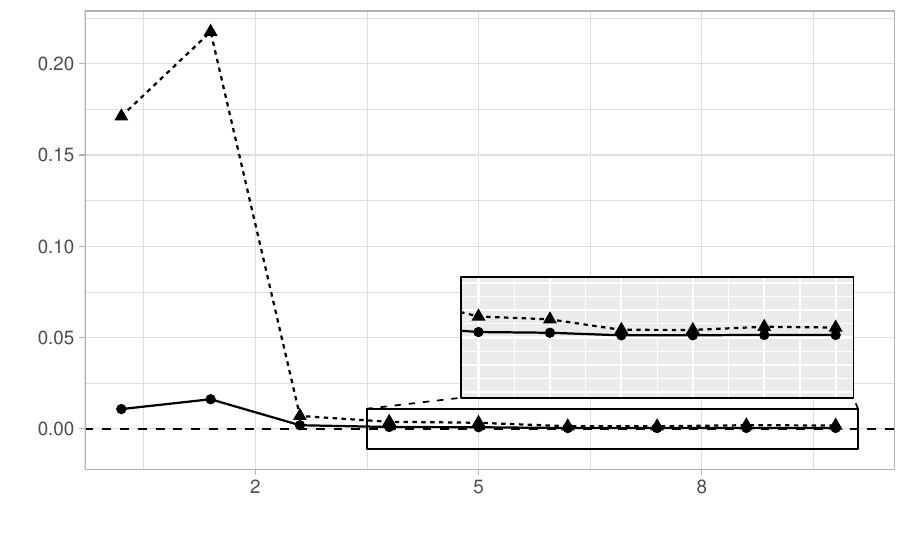}

}

}

\subcaption{\label{fig-MseSimpleAreasComp-6}Betti et al. (2006)}
\end{minipage}
\newline

\begin{minipage}[t]{.3\textwidth}

{\centering 

\raisebox{-\height}{

\includegraphics[width=\textwidth]{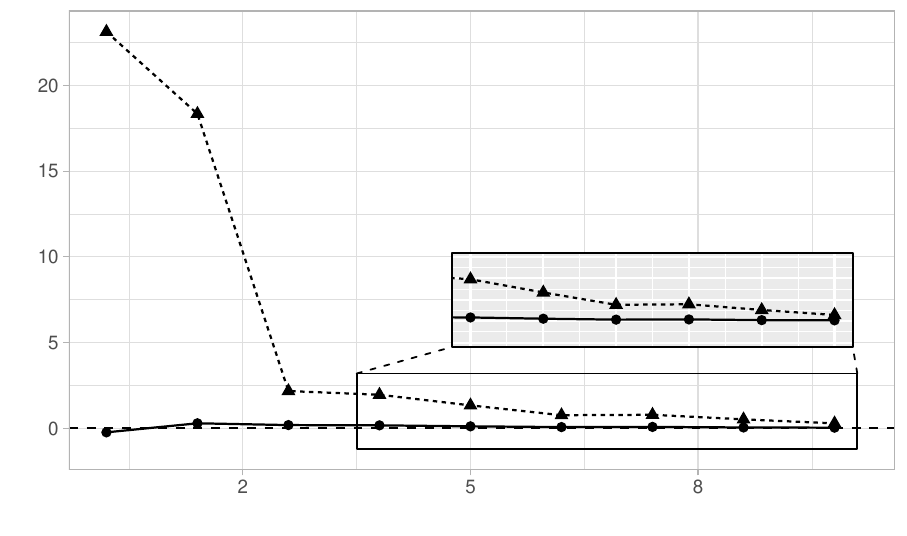}

}

}

\subcaption{\label{fig-MseSimpleAreasCompZBM}Zedini and Belhadj (2015)}
\end{minipage}%

\caption{\label{fig-MseSimpleAreasComp}Complex simulation: Bias MSE estimator for each  index at area
level $\bullet$ bootstrap and $\blacktriangle$ jackknife. Areas sorted by increasing sample size}

\end{figure}

Eventually, we investigate empirically whether one of the two estimators considered (i.e. 
jackknife or bootstrap) is to be preferred over the other. Table \ref{tab4} reports the values of
ATMSE, AEMSE and ABMSE for every index and for each estimator. Figure \ref{fig-MseSimpleAreasComp} 
reports the Bias of the MSE estimators for each area. 
It is evident that with a sample size higher than 50 the two estimators performs almost
identically for every index. The differences between the two are,
therefore, to be found for those areas where the number is very low. In this situation,
the bootstrap seems to perform better than the jackknife.

\begin{table}[H]
\footnotesize
\centering
\begin{tabular}{p{0.5cm}p{1.5cm}p{1.5cm}p{1.5cm}p{1.5cm}p{1.5cm}p{1.5cm}p{1.5cm}}  
\hline
$n_i$ & Belhadj (2014)  & Cerioli and Zani (1999) & Chacravarty (2019) & Cheli and Lemmi (1995) & Betti and Verma (1999) & Betti et. al (2006) & Zedini and Belhadj (2015)  \\  
 \hline
\multicolumn{8}{c}{ATMSE}\\
  \hline
 
3 & 4.186E-05 & 1.315E-04 & 4.753E-05 & 4.171E-04 & 4.188E-04 & 3.989E-04 & 9.810E-01 \\ 
6 & 1.051E-03 & 2.930E-04 & 1.692E-03 & 4.735E-04 & 4.847E-04 & 6.148E-04 & 4.228E-01 \\ 
48 & 4.303E-07 & 1.923E-05 & 4.351E-06 & 1.448E-05 & 1.426E-05 & 1.133E-05 & 5.473E-02 \\ 
55 & 1.849E-06 & 4.450E-05 & 1.770E-05 & 8.059E-05 & 8.133E-05 & 9.018E-05 & 1.254E-02 \\ 
100 & 1.347E-07 & 3.524E-05 & 3.374E-06 & 4.345E-06 & 4.054E-06 & 1.740E-06 & 1.829E-02 \\ 
124 & 2.862E-08 & 3.574E-05 & 9.058E-08 & 1.526E-05 & 1.507E-05 & 1.334E-05 & 7.606E-05 \\ 
128 & 1.799E-06 & 5.899E-05 & 6.247E-06 & 3.711E-05 & 3.747E-05 & 4.096E-05 & 6.018E-04 \\ 
157 & 2.429E-05 & 5.453E-07 & 1.984E-05 & 8.905E-08 & 1.048E-07 & 1.032E-07 & 7.037E-03 \\ 
187 & 1.860E-05 & 3.279E-06 & 8.294E-06 & 5.851E-06 & 5.827E-06 & 6.439E-06 & 4.127E-04 \\
808 & 8.078E-06 & 9.161E-06 & 1.720E-06 & 1.079E-05 & 1.079E-05 & 1.079E-05 & 1.021E-04 \\
   \hline
\multicolumn{8}{c}{AEMSE-Bootstrap}\\
  \hline
3 & 5.563E-03 & 1.011E-02 & 8.446E-03 & 8.914E-03 & 9.148E-03 & 1.122E-02 & 7.467E-01 \\ 
6 & 7.901E-03 & 8.194E-03 & 1.084E-02 & 1.316E-02 & 1.339E-02 & 1.682E-02 & 7.176E-01 \\ 
48 & 8.008E-04 & 1.269E-03 & 1.434E-03 & 1.608E-03 & 1.629E-03 & 1.985E-03 & 2.454E-01 \\ 
55 & 3.936E-04 & 8.117E-04 & 7.921E-04 & 9.101E-04 & 9.228E-04 & 1.096E-03 & 1.872E-01 \\ 
100 & 3.374E-04 & 6.618E-04 & 6.164E-04 & 7.227E-04 & 7.354E-04 & 8.768E-04 & 1.351E-01 \\ 
124 & 1.591E-04 & 3.512E-04 & 3.152E-04 & 3.491E-04 & 3.541E-04 & 4.179E-04 & 7.326E-02 \\ 
128 & 1.856E-04 & 3.837E-04 & 3.433E-04 & 3.799E-04 & 3.857E-04 & 4.539E-04 & 8.362E-02 \\ 
157 & 2.015E-04 & 3.083E-04 & 3.610E-04 & 3.909E-04 & 3.988E-04 & 4.803E-04 & 5.692E-02 \\ 
187 & 2.192E-04 & 3.250E-04 & 3.703E-04 & 3.828E-04 & 3.907E-04 & 4.774E-04 & 3.965E-02 \\ 
808 & 5.431E-05 &  9.437E-05 & 9.381E-05 & 5.342E-06 & 5.000E-06 & 2.271E-05 & 1.901E-04 \\
    \hline
\multicolumn{8}{c}{AEMSE-Jackknife}\\
  \hline
3 & 1.027E-01 & 1.765E-01 & 1.569E-01 & 1.314E-01 & 1.347E-01 & 1.715E-01 & 2.412E+01 \\ 
6 & 1.123E-01 & 1.301E-01 & 1.632E-01 & 1.643E-01 & 1.688E-01 & 2.181E-01 & 1.877E+01 \\ 
48 & 2.615E-03 & 4.852E-03 & 4.952E-03 & 5.867E-03 & 5.968E-03 & 7.065E-03 & 2.231E+00 \\ 
55 & 1.143E-03 & 2.988E-03 & 2.696E-03 & 3.381E-03 & 3.427E-03 & 3.933E-03 & 1.972E+00 \\ 
100& 1.030E-03 & 2.353E-03 & 2.105E-03 & 2.865E-03 & 2.898E-03 & 3.327E-03 & 1.359E+00 \\ 
124 & 4.177E-04 & 1.215E-03 & 8.922E-04 & 1.362E-03 & 1.361E-03 & 1.423E-03 & 7.664E-01 \\ 
128 & 4.803E-04 & 1.365E-03 & 9.663E-04 & 1.338E-03 & 1.341E-03 & 1.433E-03 & 7.907E-01 \\ 
157 & 5.996E-04 & 1.022E-03 & 1.176E-03 & 1.780E-03 & 1.786E-03 & 1.968E-03 & 5.312E-01 \\ 
187 & 6.472E-04 & 8.861E-04 & 1.072E-03 & 1.683E-03 & 1.685E-03 & 1.835E-03 & 2.906E-01 \\ 
808 & 0.0004 & 0.0006 & 0.0006 & 1.5440E-05&  1.2858E-05 &  1.2583E-05 & 2.002E-04\\
  \hline
 \multicolumn{8}{c}{ABMSE-Bootstrap}\\
   \hline
3& 0.0055 & 0.0100 & 0.0084 & 0.0085 & 0.0087 & 0.0108 & -0.2343 \\ 
6 & 0.0068 & 0.0079 & 0.0092 & 0.0127 & 0.0129 & 0.0162 & 0.2948 \\ 
48 & 0.0008 & 0.0012 & 0.0014 & 0.0016 & 0.0016 & 0.0020 & 0.1906 \\ 
55& 0.0004 & 0.0008 & 0.0008 & 0.0008 & 0.0008 & 0.0010 & 0.1747 \\ 
100 & 0.0003 & 0.0006 & 0.0006 & 0.0007 & 0.0007 & 0.0009 & 0.1168 \\ 
124 & 0.0002 & 0.0003 & 0.0003 & 0.0003 & 0.0003 & 0.0004 & 0.0732 \\ 
128 & 0.0002 & 0.0003 & 0.0003 & 0.0003 & 0.0003 & 0.0004 & 0.0830 \\ 
157 & 0.0002 & 0.0003 & 0.0003 & 0.0004 & 0.0004 & 0.0005 & 0.0499 \\ 
187 & 0.0002 & 0.0003 & 0.0004 & 0.0004 & 0.0004 & 0.0005 & 0.0392 \\ 
808 & 4.624E-05 & 8.521E-05 & 9.209E-05 & -5.454E-06 & -5.794E-06 & 1.192E-05 & 1.010E-06\\
   \hline
 \multicolumn{8}{c}{ABMSE-Jackkife}\\
   \hline

3 & 0.1027 & 0.1764 & 0.1569 & 0.1309 & 0.1342 & 0.1711 & 23.1356 \\ 
  6 & 0.1112 & 0.1299 & 0.1615 & 0.1638 & 0.1683 & 0.2175 & 18.3474 \\ 
  48 & 0.0026 & 0.0048 & 0.0049 & 0.0059 & 0.0060 & 0.0071 & 2.1764 \\ 
  55 & 0.0011 & 0.0029 & 0.0027 & 0.0033 & 0.0033 & 0.0038 & 1.9591 \\ 
  100 & 0.0010 & 0.0023 & 0.0021 & 0.0029 & 0.0029 & 0.0033 & 1.3403 \\ 
  124 & 0.0004 & 0.0012 & 0.0009 & 0.0013 & 0.0013 & 0.0014 & 0.7664 \\ 
  128 & 0.0005 & 0.0013 & 0.0010 & 0.0013 & 0.0013 & 0.0014 & 0.7901 \\ 
  157 & 0.0006 & 0.0010 & 0.0012 & 0.0018 & 0.0018 & 0.0020 & 0.5242 \\ 
  187 & 0.0006 & 0.0009 & 0.0011 & 0.0017 & 0.0017 & 0.0018 & 0.2902 \\ 
  808 & 0.0004 & 0.0006 & 0.0006 & 1.5440E-05 & 1.285E-05 & 1.241Ee-05 & 1.102E-05 \\
  \hline
\end{tabular}
\caption{\label{tab4}Complex simulation: ATMSE, AEMSE and ATMSE distinguished for bootstrap and jackknife estimates.}
\end{table}

In those specific areas both the estimators tend, to overestimate
the variance but the bootstrap performs better than the jackknife.
The bootstrap is in-fact able to reduce almost to 0 the bias of the MSE also
when the sample size is lower than 10.  It is important to note that
 low sample size sub-domains should always be considered in this type of
analysis. In fact, going back to what has been said above and to the results in 
Figure~\ref{fig-CvSimpleAreasCompl} we conclude that
the bootstrap estimator of the MSE is to be preferred over the jackknife
when the sample size is very low although for a size $<10$
the estimates remain un-publishable (according with \citep{Canada} ) and
specific small area estimation models should be considered.


\subsection{Computational times}

Here we will compare the effect of the two design over the obtained estimates comparing computational times. As shown previously the non-parametric bootstrap seems to perform better than the jackknife and this is also confirmed by the computational times. Table \ref{tab5}, in-fact, reports the computational times for both the designs. In particular reports the mean time needed to estimate MSE over the simulations' cycles. It is also reported the mean time useful to compute the national the MSE without considering sub-areas, that is, without requiring the algorithm to the split the population.  

Times shows as, both the MSE estimator, are quite fast. Bootstrap is always faster than the jackknife, as expected. The distribution based estimator and the Zedini and
Belhadj (2015) one are quite slower than the other because at each single loop of the MSE estimator need to rearrange the sample to compute the probability density function (distribution based) or to perform a bootstrap to obtain the value of a,b and c (Zedini and
Belhadj, 2015). If sub-areas are not consider the computational times are considerably lower. Given the previous considerations and the computational times we believe that the the bootstrap is to be preferred to the jackknife in every situation. 

\begin{table}[H]
\centering
\footnotesize
\begin{tabular}{p{1.5cm}p{1.5cm}p{1.5cm}p{1.5cm}p{1.5cm}p{1.5cm}p{1.5cm}p{1.5cm}}   
  \hline
& Belhadj (2014)  & Cerioli and Zani (1999) & Chacravarty (2019) & Cheli and Lemmi (1995) & Betti and Verma (1999) & Betti et. al (2006) & Zedini and Belhadj (2015)  \\   \hline
\multicolumn{8}{c}{Simple random sampling - Bootstrap }\\
 \hline
Areas & $\simeq 2.00 $ & $\simeq 1.71$ & $\simeq 1.77$ & $\simeq 26.00$ & $\simeq 41.79$ & $\simeq 33.96$ & $\simeq 187.96$ \\
National & $\simeq 0.56 $ & $\simeq 0.33$ & $\simeq0.38$ & $\simeq 18.96$ & $\simeq 39.32$ & $\simeq 31.14$ & $\simeq 165.36$ \\
 \hline
\multicolumn{8}{c}{Complex random sampling - Bootstrap }\\
  \hline
Areas &  $\simeq 1.90 $ & $\simeq 1.47$ & $\simeq 1.45$ & $\simeq 11.39$ & $\simeq 9.30$ & $\simeq 9.69$ & $\simeq 45.85$ \\
National  & $\simeq 0.38 $ & $\simeq 0.20$ & $\simeq 0.18$ & $\simeq 7.36$ & $\simeq 11.06$ & $\simeq 9.22$ & $\simeq 23.30$ \\
 \hline
\multicolumn{8}{c}{Complex random sampling - Jackknife }\\
  \hline

  Areas & $\simeq 0.45 $ & $\simeq 0.41$ & $\simeq 0.41$ & $\simeq 98.88 $ & $\simeq 134.99$ & $\simeq 115.45$ & $\simeq 173.90$ \\
National & $\simeq 0.23$ & $\simeq 0.21$ & $\simeq 0.19$ & $\simeq 88.80$ & $\simeq 129.15$ & $\simeq 112.64$ & $\simeq  164.51$ \\
\hline
\end{tabular}
\caption{\label{tab5}Computational times (Seconds)}
\end{table}

\section{Robustness analysis}\label{sec-robusteness}

In this section, we analyze the robustness of the normative indices.
The reader may have noticed that some fuzzy indices depends on certain
parameters (i.e. $z_1$, $z_2$, $\beta$) that need to be specified. 
With this statement we refer to a choice of parameters that
it is not estimated from data or set a-priori.

Therefore, to decide which fuzzy index is preferable or usable to report
is non-trivial. One should choose the index that is most robust, meaning
the one that is least sensitive to changes in the sources of
uncertainty. To have a reliable and consistent 
index, one should undertake robustness analysis followed by classical statistical
inference. Robustness analysis involves checking the ordering of two or
more areas under alternative choices over the parameters specification
of the index. In particular, we focus the indices that has a membership function
defined as normative with the aims to point out the role played by the
arbitrary constant. The robustness check is done at national level
and at the area level. 

At national level, the MSE of the indices is analyzed, at the national level, to
figure out if in what extent the choice of the parameter can influence the
variability. In other words, if the quality of the estimates,
  according to the guidelines of \citep{Canada} changes
  with respect to the chosen parameters. 
  
At area level, the order of the estimates are checked with the methods of
  ``correlation coefficients for ranks'' (\citep{alkire2018new}). These methods test if an
  ordering of more than two areas remains the same when the value of
  some parameter is altered. The robustness of a ranking is evaluated by
  calculating a rank correlation coefficient between the initial and the
  alternative rankings. In this section the reported results are only those regarding the Complex simulation (results for the SRS are almost identical and available under request).

\hypertarget{are-the-coefficients-influent-on-the-mse-value}{%
\subsection{Are the coefficients influent on the MSE
value?}\label{are-the-coefficients-influent-on-the-mse-value}}

\begin{figure}

\begin{minipage}[t]{0.50\linewidth}

{\centering 

\raisebox{-\height}{

\includegraphics{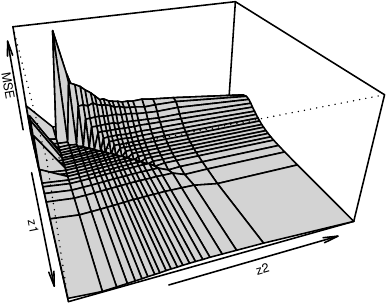}

}

}

\subcaption{\label{fig-MSEbelhadj-1}$\beta$ = 1}
\end{minipage}%
\begin{minipage}[t]{0.50\linewidth}

{\centering 

\raisebox{-\height}{

\includegraphics{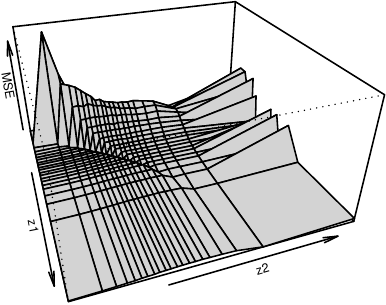}

}

}

\subcaption{\label{fig-MSEbelhadj-2}$\beta$ = 2}
\end{minipage}%
\newline
\begin{minipage}[t]{0.50\linewidth}

{\centering 

\raisebox{-\height}{

\includegraphics{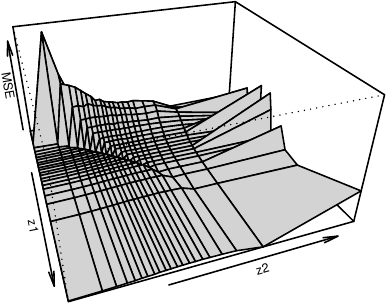}

}

}

\subcaption{\label{fig-MSEbelhadj-3}$\beta$ = 4}
\end{minipage}%
\begin{minipage}[t]{0.50\linewidth}

{\centering 

\raisebox{-\height}{

\includegraphics{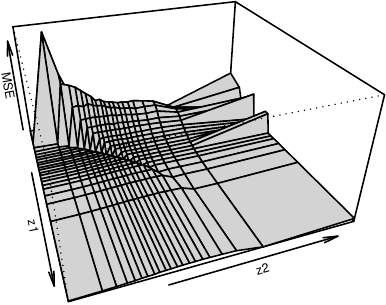}

}

}

\subcaption{\label{fig-MSEbelhadj-4}$\beta$ = 10}
\end{minipage}%

\caption{\label{fig-MSEbelhadj}MSE of Belhadj (2014).}

\end{figure}

Figure~\ref{fig-MSEbelhadj} shows clearly that the MSE of
the sample average of the membership function in
Equation~\ref{eq-belhadj2015} is sensitive to the choices of the
thresholds. In particular, setting high values of $z_1$ will usually
decrease the variance of the estimator while the opposite happens for
small values. Note that from Equation~\ref{eq-belhadj2015} the portion
of the plane $(z_1,z_2)$ to be considered is that of the points
$\{(z_1,z_2): z_2 \ge z_1\}$. Regarding the role of the parameter $\beta$, it seems to take
the role of inflating the variance is the upper region of the plane.

The same considerations apply to Equation~\ref{eq-cerioli} although for
this membership function the mean squared error surface has a much more
regular shape. For what regards Equation~\ref{eq-chackra} instead, the
mean squared error seems to be not dependent on the choice of the
threshold as long as the threshold is sufficiently large.

These pictures also suggest another important aspect to consider when
estimating fuzzy measure, that is, a mis-specified selection of the
thresholds that depending on the surface of the mean squared error may
result in severe biases. Suppose that $z_1^*$ and $z_2^*$ are the true values
($b = 1$ without loss of generality) and that MSE$(z_1^*, z_2^*)$ is the
corresponding MSE. Suppose that $z'_1$ and $z'_2$ are the values specified
by the researcher (again, $\beta = 1$ without loss of generality) and 
that MSE$(z'_1, z'_2)$ is the corresponding MSE. The difference between the
two MSE may be not-negligible in some portions of the $z'_1, z'_2$ plane.

\begin{figure}

{\centering \includegraphics{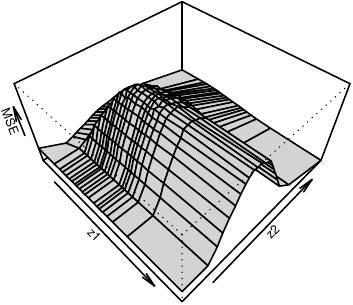}

}

\caption{\label{fig-MSECerioliSimple}MSE of Cerioli and Zani (1999)}

\end{figure}

\begin{figure}

{\centering \includegraphics{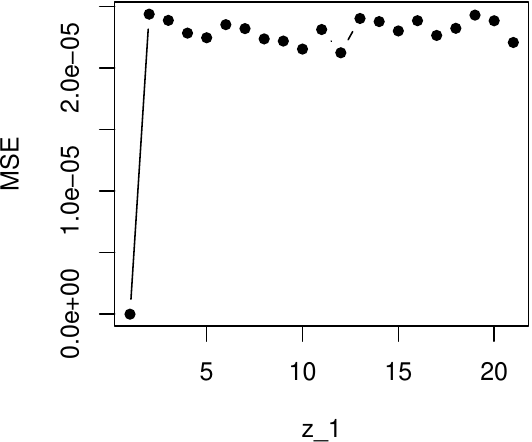}

}

\caption{\label{fig-MSEChakraSimple}MSE of Chakravarty (2019)}

\end{figure}



\hypertarget{correlation-coefficients-for-ranks}{%
\subsection{Correlation coefficients for
ranks}\label{correlation-coefficients-for-ranks}}

Suppose there are $j^*$ areas, and two set of ranks, $r$ and $r^{*}$
for two different specification sets of parameters regarded as the initial and alternative specification. 
Also, consider $r_{j^*}$ and $r^{*}_{j^*}$ as the ranks of the area $j^*$ for the two
different specification sets of parameters in the same membership function. The
alternative specification may entail a change in the set of the
$\alpha-cut$ parameters or in the $\beta$ parameters. 
According with \citep{corretest} we use the Kendal and the Spearman 
correlation coefficients. The former is useful for those situation where the sample size is small and there are many tied ranks.  Table~\ref{tabr1} reports the mean values of the two correlation coefficients for the three index. In the first table the value $\beta$ of the Belhadj (2014) index is equal to 2. For Cerioli and Zani (1999) and Chacravarty (2019) indices  we use as benchmark the indices computed with the parameters in Table~\ref{tabparam}. 
The values for both the correlation indices tends to decrease fast except for Cerioli and Zani (1999). 
In particular the Belhadj (2014) index seems to not be robust. The same conclusions arise also from Table~\ref{tabr2}. 
In this table, in-fact, the value of the correlation are computed for the Belhadj (2014) index
for fixed $z_1$ and $z_2$, according with Table \ref{tabparam}, and 
changing the value of $\beta$, again, this index seems to be not robust.

\begin{table}[H]
\centering
\footnotesize
\begin{tabular}{rrrrrrrrrrr}  
  \hline

Index & Correlation & \multicolumn{9}{c}{Parameters }\\
  \hline
&    & \makecell[c]{$Y_{0.02}$\\$Y_{0.98}$} & \makecell[c]{$Y_{0.03}$\\$Y_{0.97}$} & \makecell[c]{$Y_{0.04}$\\$Y_{0.96}$} & \makecell[c]{$Y_{0.05}$\\$Y_{0.95}$} & \makecell[c]{$Y_{0.06}$\\$Y_{0.94}$} & \makecell[c]{$Y_{0.07}$\\$Y_{0.93}$} & \makecell[c]{$Y_{0.08}$\\$Y_{0.92}$} & \makecell[c]{$Y_{0.09}$\\$Y_{0.91}$} & \makecell[c]{$Y_{0.10}$\\$Y_{0.90}$} \\
  \hline
\multirow{2}{*}{Belhadj (2014), $\beta=2$}& Kendal & 0.78 & 0.67 & 0.67 & 0.67 & 0.56 & 0.50 & 0.50 & 0.56 & 0.50 \\ 
  & Spearman & 0.90 & 0.85 & 0.85 & 0.83 & 0.73 & 0.70 & 0.70 & 0.70 & 0.67\\
  \hline
  \multirow{2}{*}{Cerioli and Zani (1999)}&  Kendal & 0.94 & 0.94 & 0.89 & 0.89 & 0.89 & 0.89 & 0.89 & 0.89 & 0.89 \\ 
 & Spearman & 0.98 & 0.98 & 0.97 & 0.97 & 0.97 & 0.97 & 0.97 & 0.97 & 0.97 \\ 
 \hline
 Index & Correlation & \multicolumn{9}{c}{Parameters}\\
 \hline
  &   & $Y_{0.30}$ & $Y_{0.35}$ & $Y_{0.40}$ & $Y_{0.45}$  & $Y_{0.55}$ &$ Y_{0.60}$ & $Y_{0.65}$ & $Y_{0.70}$ &  $Y_{0.75}$\\
  \hline
   \multirow{2}{*}{Chacravarty (2019)} &  Kendal & 0.94 & 0.89 & 0.83 & 0.78 & 0.78 & 0.72 & 0.67 & 0.67 & 0.67 \\ 
   & Spearman & 0.98 & 0.95 & 0.90 & 0.88 & 0.88 & 0.87 & 0.78 & 0.78 & 0.78 \\
   \hline
\end{tabular}
\caption{\label{tabr1}Correlation coefficients for ranks changing $z$s' parameters}
\end{table}

\begin{table}[H]
\centering
\footnotesize
\begin{tabular}{rrrrrrrrrrr}  
  \hline

Index & Correlation & \multicolumn{9}{c}{Parameters: $\beta$}\\
  \hline
&    &  $\beta=1$ & $\beta=3$ & $\beta=4$ & $\beta=5$& $\beta=6$ & $\beta=7$ & $\beta=8$ & $\beta=9$ & $\beta=10$ \\
  \hline
\multirow{2}{*}{Belhadj (2014)}& Kendal &0.91&  0.89 & 0.83 & 0.78 & 0.72 & 0.72 & 0.72 & 0.67 & 0.67 \\ 
  & Spearman & 0.99 & 0.97 & 0.93 & 0.90 & 0.85 & 0.85 & 0.85 & 0.83 & 0.83\\
  \hline
  
\end{tabular}
\caption{\label{tabr2}Correlation coefficients for ranks for Belhadj index changing $\beta$}
\end{table}

\hypertarget{sec-conclusions}{%
\section{Conclusions}\label{sec-conclusions}}

This paper has addressed the issue of estimating fuzzy poverty indices.
Starting from a real sample of individuals, we simulated a synthetic
population of individuals and started sampling from that population to
address the issue of estimating uncertainty of fuzzy poverty indices. We
first investigated the bias of sampling statistics as well as the bias
of the estimation of their mean squared error by simple random sampling
of individual from the population. In order to not loose touch with
realistic survey that usually involve stratification and more complex
designs we repeated the analysis to allow also for this more realist
scenario.

We noticed that if we use the sample mean of the sampling estimate of
the membership function we obtain unbiased estimated with low
coefficient of variations. Also, for those methods that require
parameter specification we draw the surface of the mean squared error
for different values of the parameters. This has borough many points of
reflection as our simulations show that a mis-specification of the
parameters can lead to severe distortions when estimating the mean
squared errors. Also it has shown that accurate estimates of fuzzy
indices can be just the result of poor specification of parameters.

At least two points of reflection emerge from these considerations. The
first is of a theoretical kind, that is, we treated these parameters as
unknown values in the population of individuals that are fixed and
unknown to researcher. Although this is a necessary working assumption to
discuss about bias in estimation procedures, the fact that these could
be unknown parameter of a population is something that has not been
discussed extensively in the literature. The fact that these parameter
are set arbitrarily without a statistical procedure for their estimation
based on maximum livelihood or other methods make difficult to draw a
comparison without making assumptions of the kind above and on which
sampling measure to estimate.

Regarding this last point, we focused our study assuming as sample
statistics the average of the membership function. Although the mean is a
solid statistical estimator, not all the methods discussed originally
suggest this measure. The lack of consistency between the various
approaches suggested in the literature and here reported had us make the
choice of the sample mean for two reasons, one statistical, and the
second because it is the most used statistic in the methods considered.
For this reason, our result shall not be intended as an assessment of
only the original papers but considering the fact that we had to make a
common ground of comparison.

Anyways, we believe that some kind of baseline is needed in this field
of research and we hope that we made a step in this direction with this
work.



\begin{thebibliography}{32}
\providecommand{\natexlab}[1]{#1}
\providecommand{\url}[1]{\texttt{#1}}
\expandafter\ifx\csname urlstyle\endcsname\relax
  \providecommand{\doi}[1]{doi: #1}\else
  \providecommand{\doi}{doi: \begingroup \urlstyle{rm}\Url}\fi

\bibitem[Lemmi and Betti(2006)]{lemmi2006fuzzy}
Achille~A Lemmi and Gianni Betti.
\newblock \emph{Fuzzy set approach to multidimensional poverty measurement},
  volume~3.
\newblock Springer Science \& Business Media, 2006.

\bibitem[Betti and Lemmi(2008)]{betti2008advances}
Gianni Betti and Achille Lemmi.
\newblock \emph{Advances on income inequality and concentration measures},
  volume 102.
\newblock Routledge, 2008.

\bibitem[Betti and Lemmi(2021)]{betti2021analysis}
Gianni Betti and Achille Lemmi.
\newblock \emph{Analysis of Socio-Economic Conditions: Insights from a Fuzzy
  Multi-dimensional Approach}.
\newblock Routledge, 2021.

\bibitem[Kakwani(1993)]{kakwani1993statistical}
Nanak Kakwani.
\newblock Statistical inference in the measurement of poverty.
\newblock \emph{The Review of Economics and Statistics}, pages 632--639, 1993.

\bibitem[Graf and Till{\'e}(2014)]{graf2014variance}
Eric Graf and Yves Till{\'e}.
\newblock Variance estimation using linearization for poverty and social
  exclusion indicators.
\newblock \emph{Survey Methodology}, 40\penalty0 (1):\penalty0 61--80, 2014.

\bibitem[Duclos et~al.(2006)Duclos, Sahn, and Younger]{duclos2006robust}
Jean-Yves Duclos, David~E Sahn, and Stephen~D Younger.
\newblock Robust multidimensional poverty comparisons.
\newblock \emph{The economic journal}, 116\penalty0 (514):\penalty0 943--968,
  2006.

\bibitem[Alfons and Templ(2012)]{alfons2012estimation}
Andreas Alfons and Matthias Templ.
\newblock Estimation of social exclusion indicators from complex surveys: The r
  package laeken.
\newblock \emph{KU Leuven, Faculty of Business and Economics Working Paper},
  2012.

\bibitem[Betti et~al.(2018)Betti, Gagliardi, and Verma]{betti2018simplified}
Gianni Betti, Francesca Gagliardi, and Vijay Verma.
\newblock Simplified jackknife variance estimates for fuzzy measures of
  multidimensional poverty.
\newblock \emph{International Statistical Review}, 86\penalty0 (1):\penalty0
  68--86, 2018.

\bibitem[Fisher(1997)]{fisher1997development}
Gordon~M Fisher.
\newblock The development of the orshansky poverty thresholds and their
  subsequent history as the official u.s. poverty measure.
\newblock \emph{United States Census Bureau}, 1997.

\bibitem[Fisher(1992)]{fisher1992development}
Gordon~M Fisher.
\newblock The development and history of the poverty thresholds.
\newblock \emph{Soc. Sec. Bull.}, 55:\penalty0 3, 1992.

\bibitem[Allen(2013)]{allen2013poverty}
Robert Allen.
\newblock Poverty lines in history, theory, and current international practice.
\newblock 2013.

\bibitem[Gillie(1996)]{History1}
Alan Gillie.
\newblock The origin of the poverty line.
\newblock \emph{The Economic History Review}, 49\penalty0 (4):\penalty0
  715--730, 1996.

\bibitem[Rowntree(1901)]{Rowntree}
Ba~Sa Rowntree.
\newblock \emph{Poverty: A Study of Town Life}.
\newblock London,Macmillan., 1901.

\bibitem[Jolly et~al.(1976)]{jolly1976world}
Richard Jolly et~al.
\newblock The world employment conference: The enthronement of basic needs.
\newblock \emph{Development Policy Review}, 9\penalty0 (2):\penalty0 31--44,
  1976.

\bibitem[Cheli and Lemmi(1995)]{chelilemmi}
Bruno Cheli and Achille Lemmi.
\newblock A’totally’fuzzy and relative approach to the multidimensional
  analysis of poverty.
\newblock 1995.

\bibitem[Zadeh(1965)]{zadeh1965fuzzy}
Lotfi~A Zadeh.
\newblock Fuzzy sets.
\newblock \emph{Information and control}, 8\penalty0 (3):\penalty0 338--353,
  1965.

\bibitem[Klir and Yuan(1995)]{klir1995fuzzy}
George Klir and Bo~Yuan.
\newblock \emph{Fuzzy sets and fuzzy logic}, volume~4.
\newblock Prentice hall New Jersey, 1995.

\bibitem[Martinetti(2006)]{martinetti2006capability}
Enrica~Chiappero Martinetti.
\newblock Capability approach and fuzzy set theory: description, aggregation
  and inference issues.
\newblock \emph{Fuzzy set approach to multidimensional poverty measurement},
  pages 93--113, 2006.

\bibitem[Cerioli and Zani(1990)]{cerioli1990fuzzyart}
Andrea Cerioli and Sergio Zani.
\newblock A fuzzy approach to the measurement of poverty.
\newblock In \emph{Income and Wealth Distribution, Inequality and Poverty:
  Proceedings of the Second International Conference on Income Distribution by
  Size: Generation, Distribution, Measurement and Applications, Held at the
  University of Pavia, Italy, September 28--30, 1989}, pages 272--284.
  Springer, 1990.

\bibitem[Belhadj(2011)]{belhadj2011new}
Besma Belhadj.
\newblock A new fuzzy unidimensional poverty index from an information theory
  perspective.
\newblock \emph{Empirical Economics}, 40:\penalty0 687--704, 2011.

\bibitem[Zedini and Belhadj(2015)]{zedini2015new}
Asma Zedini and Besma Belhadj.
\newblock A new approach to unidimensional poverty analysis: Application to the
  t unisian case.
\newblock \emph{Review of Income and Wealth}, 61\penalty0 (3):\penalty0
  465--476, 2015.

\bibitem[Belhadj(2014)]{besma2014employment}
Besma Belhadj.
\newblock Employment measure in developing countries via minimum wage and
  poverty new fuzzy approach.
\newblock \emph{Opsearch}, 52:\penalty0 329--339, 2014.

\bibitem[Betti and Verma(1999)]{betti1999measuring}
Gianni Betti and Vijay Verma.
\newblock Measuring the degree of poverty in a dynamic and comparative context:
  a multi-dimensional approach using fuzzy set theory.
\newblock In \emph{Proceedings, iccs-vi}, volume~11, page 289, 1999.

\bibitem[Betti et~al.(2006)Betti, Cheli, Lemmi, and
  Verma]{betti2006multidimensional}
Gianni Betti, Bruno Cheli, Achille Lemmi, and Vijay Verma.
\newblock Multidimensional and longitudinal poverty: an integrated fuzzy
  approach.
\newblock \emph{Fuzzy set approach to multidimensional poverty measurement},
  pages 115--137, 2006.

\bibitem[Verma and Betti(2011)]{verma2011taylor}
Vijay Verma and Gianni Betti.
\newblock Taylor linearization sampling errors and design effects for poverty
  measures and other complex statistics.
\newblock \emph{Journal of Applied Statistics}, 38\penalty0 (8):\penalty0
  1549--1576, 2011.

\bibitem[Berger and Skinner(2005)]{berger2005jackknife}
Yves~G Berger and Chris~J Skinner.
\newblock A jackknife variance estimator for unequal probability sampling.
\newblock \emph{Journal of the Royal Statistical Society Series B: Statistical
  Methodology}, 67\penalty0 (1):\penalty0 79--89, 2005.

\bibitem[Templ et~al.(2017)Templ, Meindl, Kowarik, and
  Dupriez]{templ2017simulation}
Matthias Templ, Bernhard Meindl, Alexander Kowarik, and Olivier Dupriez.
\newblock Simulation of synthetic complex data: The r package simpop.
\newblock \emph{Journal of Statistical Software}, 79\penalty0 (10):\penalty0
  1--38, 2017.

\bibitem[Deming and Stephan(1940)]{deming1940least}
W~Edwards Deming and Frederick~F Stephan.
\newblock On a least squares adjustment of a sampled frequency table when the
  expected marginal totals are known.
\newblock \emph{The Annals of Mathematical Statistics}, 11\penalty0
  (4):\penalty0 427--444, 1940.

\bibitem[Kv{\aa}lseth(2017)]{kvaalseth2017coefficient}
Tarald~O Kv{\aa}lseth.
\newblock Coefficient of variation: the second-order alternative.
\newblock \emph{Journal of Applied Statistics}, 44\penalty0 (3):\penalty0
  402--415, 2017.

\bibitem[Statistics-Canada(2007)]{Canada}
Statistics-Canada.
\newblock 2005 survey of financial security - public use microdata file, user
  guide. published by authority of the minister responsible for statistics
  canada.
\newblock Technical report, 2007.

\bibitem[Alkire and Jahan(2018)]{alkire2018new}
Sabina Alkire and Selim Jahan.
\newblock The new global mpi 2018: Aligning with the sustainable development
  goals.
\newblock 2018.

\bibitem[Noha and Sama(2021)]{corretest}
Omar Noha and El~Hage~Sleiman. Sama.
\newblock Testing the robustness of the revised multidimensional poverty index
  for arab countries.
\newblock \emph{United Nation Bureau Economic and Social Commission for Western
  Asia}, 2021.

\end{thebibliography}
\end{document}